\documentclass[10pt,a4paper]{article}
\usepackage{jcappub}
\allowdisplaybreaks

\usepackage{ifthen}
\usepackage{graphicx}
\usepackage{dcolumn}
\usepackage{bm}
\usepackage{color}
\usepackage{amsmath}
\usepackage{amssymb}
\usepackage{subfigure}

\newcommand{\vp}{\mathbf{p}}
\newcommand{\vr}{\mathbf{r}}
\newcommand{\vs}{\mathbf{s}}

\newcommand{\vx}{\mathbf{x}}

\newcommand{\vk}{\mathbf{k}}
\newcommand{\vz}{\mathbf{z}}

\newcommand{\vv}{\mathbf{v}}

\newcommand{\himpc}{{\hbox {$~h^{-1}$}{\rm ~Mpc}}}
\newcommand{\hmpci}{{\hbox {$~h{\rm ~Mpc}^{-1}$}}}

\newcommand{\be}{\begin{equation}}
\newcommand{\ee}{\end{equation}}
\newcommand{\bey}{\begin{eqnarray}}
\newcommand{\eey}{\end{eqnarray}}

\newcommand*{\VEC}[1]  {\textbf{#1}}
\newcommand*{\pp}  {\parallel}

\newcommand*{\df}  {\delta}

\begin{document}

\title{
Peculiar velocities in redshift space: formalism, $N$-body simulations and perturbation theory}

\author[a,b]{Teppei Okumura,} \emailAdd{teppei.okumura@ipmu.jp}
\author[a,c,d]{Uro{\v s} Seljak,} \emailAdd{useljak@berkeley.edu}
\author[d]{Zvonimir Vlah,} \emailAdd{zvlah@physik.uzh.ch}
\author[e]{and Vincent Desjacques} \emailAdd{Vincent.Desjacques@unige.ch}

\affiliation[a]{Institute for the Early Universe, Ewha Womans
  University, Seoul 120-750, S. Korea}
  
\affiliation[b]{Kavli Institute for the Physics and Mathematics of the Universe (Kavli IPMU, WPI), The University of Tokyo, Chiba 277-8582, Japan}
  
\affiliation[c]{Department of Physics, Department of Astronomy, and Lawrence Berkeley National
  Laboratory, University of California, Berkeley, California 94720,
  USA} 

\affiliation[d]{Physik Institut, University of Zurich, 8057 Zurich, Switzerland}

\affiliation[e]{D\'epartement de Physique Th\'eorique and Center for Astroparticle Physics (CAP), 
Universit\'e de Gen\`eve, 1211 Gen\`eve, Switzerland}

\abstract{
Direct measurements of peculiar velocities of galaxies and clusters of galaxies can in principle provide explicit information on the three dimensional mass distribution, but this information is modulated by the fact that velocity field is sampled at galaxy positions, and is thus probing galaxy momentum. We derive expressions for the cross power spectrum between the density and momentum field and the auto spectrum of the momentum field in redshift space, by extending the distribution function method to these statistics. The resulting momentum cross and auto power spectra in redshift space are expressed as infinite sums over velocity moment correlators in real space, as is the case for the density power spectrum in redshift space. We compute each correlator using Eulerian perturbation theory (PT) and halo biasing model and compare the resulting redshift-space velocity statistics to those measured from $N$-body simulations for both dark matter and halos. We find that in redshift space linear theory predictions for the density-momentum cross power spectrum as well as for the momentum auto spectrum fail to predict the $N$-body results at very large scales. On the other hand, our nonlinear PT prediction for these velocity statistics, together with real-space power spectrum for dark matter from simulations, improves the accuracy for both dark matter and halos. We also present the same analysis in configuration space, computing the redshift-space pairwise mean infall velocities and velocity correlation function and compare to nonlinear PT. 
}
\keywords{galaxy clustering, power spectrum, redshift surveys, cosmological perturbation theory}
\arxivnumber{1312.4214}

\maketitle

\section{Introduction}\label{sec:intro}
The peculiar velocity field probed by galaxies and clusters of galaxies is cosmologically very important because it provides information on the three dimensional mass density fluctuations. There are two observational ways to extract the information. The first one is a galaxy redshift survey, where peculiar velocities induce redshift space distortions along the line of sight (RSD) \cite{Jackson:1972, Sargent:1977, Kaiser:1987,Hamilton:1998}. The information on peculiar velocities is entangled in the clustering throuh the dependence on the line of sight angle and, thus, can be separated out from the intrinsic clustering of galaxies. The second is a direct map of peculiar velocities \cite{Davis:1977, Peebles:1980, Strauss:1995}, which can be obtained by measuring the intrinsic distances to galaxies (e.g. by using Tully-Fisher or Faber-Jackson relations) and comparing them to the galaxy redshift. Unlike redshift surveys, peculiar velocity surveys  furnish in principle a direct measurement of peculiar velocities. However, the error on the velocity measurement is proportional to the distance, so the analysis is limited to the nearby universe ($z\sim 0$). The two types of surveys are largely complementary to each other, and combining these two probes can tighten constraints on the growth rate of cosmic structure and, therefore, on theories of modified gravity \cite{Hudson:2012}. Observational analyses in peculiar velocity surveys have been performed by \cite[e.g.,][]{Freudling:1999, Juszkiewicz:2000, Park:2006, Watkins:2007} (see also \cite{Davis:2011} for recent work). More recently, it has been suggested that peculiar velocity measurements can be extended to higher redshifts using Type Ia supernovae \cite{Haugbolle:2007, Bhattacharya:2011, Turnbull:2012} or the kinetic Sunyaev-Zeldovich (kSZ, \cite{Sunyaev:1980}) effect \cite{Bhattacharya:2008}. The kSZ effect has been detected by \cite{Hand:2012} using high-resolution microwave sky maps and a large galaxy redshift survey.

Theoretical studies have attempted to predict peculiar velocity statistics using linear theory \cite{Davis:1977, Peebles:1980, Fisher:1995, Hu:2000, Park:2000, Ma:2002}, halo models \cite{Bhattacharya:2008, Lam:2013}, nonlinear PT \cite{Scoccimarro:2004, Smith:2008, Reid:2011a, Kitaura:2013} or empirical fitting functions \cite{Zu:2012}. All the above theoretical works were formulated in real space. In reality, what is measured in peculiar velocity surveys is the velocity field sampled in redshift space, and the measured positions are distorted by RSD. Theoretically, the RSD effect on peculiar velocities has been discussed previously in Fourier space statistics \cite{Burkey:2004, Ho:2009, Shao:2011, Koda:2013}. 

In this paper, we present a comprehensive analysis of the aforementioned issues for two peculiar velocity statistics: the velocity-density correlation, which can be restated as the pairwise mean infall velocity, and the velocity-velocity correlation. We have developed and tested a phase space distribution function approach to RSD where the redshift-space density can be written as a sum over mass or number weighted moments of radial velocity \cite{Seljak:2011, Okumura:2012, Okumura:2012b, Vlah:2012, Vlah:2013}. The corresponding RSD power spectrum can be written as a sum over auto and cross-correlators of these moments. By applying this technique to the momentum density field in redshift space, we derive theoretical predictions for momentum power spectra in redshift space as a sum over auto and cross correlators of density-weighted velocity moments. We present detailed comparisons of these predictions computed using Eulerian nonlinear perturbation \cite{Vlah:2012, Vlah:2013} to the measurements for dark matter and halos obtained from $N$-body simulations. 

This paper is organized as follows. In section \ref{sec:theory}, we briefly review peculiar velocity statistics in real space, and derive expressions in redshift space by extending the distribution function approach for the redshift-space density field to the redshift-space momentum field. Section \ref{sec:sim} describes the $N$-body simulations as well as our measurements of the Fourier space, peculiar velocity statistics considered in this paper. In section \ref{sec:analysis}, we present the peculiar velocity statistics in Fourier space, the density-momentum and momentum-momentum power spectra measured from $N$-body simulations. We then make a detailed comparison with linear and nonlinear perturbation theory. In section \ref{sec:configuration_analysis} we present the peculiar velocity statistics in configuration space, the mean pairwise velocity and the velocity-velocity correlation function. We then discuss the configuration-space peculiar velocity statistics measured from $N$-body simulations and predicted using perturbation theory. In section \ref{sec:satellite}, we test whether our nonlinear perturbation theory for halos can also model the peculiar velocity statistics of satellite galaxies. Section \ref{sec:conclusion} is the conclusion of this paper. For a comparison to the redshift-space velocity statistics we discuss in this paper, we present the corresponding real-space statistics in appendix \ref{sec:real_space}. In the analysis of velocity statistics in configuration space, statistics using pair-weighted velocity is commonly used while we focus on the statistics using density-weighted velocity throughout this paper. 
In appendix \ref{sec:pair_velocity}, we perform a detail comparison of pair-weighted and density-weighted statistics and show that the latter are better defined and more direct observables.

\section{Velocity moments power spectra in real and redshift space}\label{sec:theory}

In this section we briefly review theoretical formalisms for the velocity statistics in real space, and also 
newly define the statistics in redshift space.
Throughout this paper we adopt a plane-parallel approximation.
In what follows the super/subscript $m$ and $h$ denote quantities for dark matter and halos, respectively. 
We will, however, omit the $m$ and $h$ when a given equation holds for both dark matter and halos, 
because much of our formalism remains unchanged if the dark matter particles are 
replaced with biased tracers such as halos or galaxies.

In peculiar velocity surveys, peculiar velocities of galaxies themselves are directly observable, but 
they are sampled at galaxy positions, so they are not volume-weighted but density-weighted velocities. We can 
define the comoving momentum density field as 
${\bf p}(\vx) = \left[ 1+\delta (\vx)\right] \vv (\vx)$, 
where $\delta = \rho/\bar{\rho}-1$ is an overdensity field and $\vv$ is the comoving velocity field. Since only the radial components of these vectors are observables, a more relevant relation is
\be
p_{\parallel}(\vx) = \left[ 1+\delta (\vx)\right] v_\parallel (\vx),  \label{eq:momentum}
\ee
where $p_\parallel=\vp\cdot\hat{\bf z}$ and $\hat{\bf z}$ is the unit vector pointing along the line of sight. 
Equation (\ref{eq:momentum}) is the $L=1$ case of the moments of mass-weighted velocities $T_\parallel^L$ defined in \cite{Seljak:2011} (see section \ref{sec:dfa} below).

\subsection{Momentum power spectra in real space}
By correlating the radial momentum density to the density field and to the radial momentum density itself, respectively, we obtain the first-order $P_{01}^r$ and second-order $P_{11}^r$ velocity statistics, as
\bey
\left\langle\df(\VEC{k})| p^{*}_\pp(\VEC{k}')\right\rangle=(2\pi)^3 P_{01}^r(\VEC{k})\df^D(\VEC{k}-\VEC{k}'), 
\label{eq:momentum_cros_real}\\
\left\langle p_\pp(\VEC{k})| p^{*}_\pp(\VEC{k}')\right\rangle=(2\pi)^3 P_{11}^r(\VEC{k})\df^D(\VEC{k}-\VEC{k}'),
\label{eq:momentum_auto_real}
\eey
where $\delta(\vk)$ and $p_\pp(\vk)$ are the Fourier transforms of $\delta(\vx)$ and $p_\pp(\vx)$, respectively. 
Quantities with the superscript $r$ mean that they are measured in real space, while later we will consider these quantities in redshift space, denoted by the superscript $s$. The general definition of $P_{LL'}^r$ was introduced in \cite{Seljak:2011} and will be given in the next subsection. 

The momentum field is related to the density field through the continuity equation, and in linear theory for biased objects it reads
\be
P^{r,h}_{00, {\rm lin}}(\vk) = \left( \frac{-ikb}{{\cal H}\mu f} \right)P^{r,h}_{01, {\rm lin}}(\vk) = \left( \frac{kb}{{\cal H}\mu f} \right)^2 P^{r,h}_{11, {\rm lin}}(\vk), \label{eq:pll_kaiser}
\ee
where $P_{00}^r$ is the ordinary power spectrum of a given density field, ${\cal H}(a)=aH(a)$, $H(a)$ is the Hubble parameter, $\mu$ is the angle between the Fourier mode and line-of-sight, $b$ is the bias parameter and $f=d\ln{D}/d\ln{a}$ with $D$ the growth factor. The expression for dark matter is obtained by substituting $b=1$ in equation (\ref{eq:pll_kaiser}), thus the relations between power spectra for dark matter and halos in linear theory are characterized by $b$ as $P^{r,h}_{00,{\rm lin}}=b^2 P^{r,m}_{00,{\rm lin}}$, 
$P^{r,h}_{01,{\rm lin}}=b P^{r,m}_{01,{\rm lin}}$ and $P^{r,h}_{11,{\rm lin}}=P^{r,m}_{11,{\rm lin}}$ \cite{Seljak:2011, Okumura:2012b}.


\subsection{Density power spectrum from the distribution function approach}\label{sec:dfa}
A phase-space distribution function approach was recently developed by \cite{McDonald:2011, Seljak:2011} for the purpose of modeling the power spectrum of the density fields in redshift-space. 
The approach was tested against dark matter in \cite{Okumura:2012, Vlah:2012} and extended to biased objects 
such as dark matter halos and galaxies in \cite{Okumura:2012b, Vlah:2013}.
In this subsection we briefly review the formalism.

The redshift-space density field in Fourier space is described as a summation of moments of density-weighted 
velocity in real space,
\be
\delta^s(\vk)=\sum_{L=0}\frac{1}{L!}\left(\frac{ik_\parallel}{\cal H}\right)^L T_\parallel^{L}(\vk). \label{eq:delta_k}
\ee
$T_{\parallel}^{L}(\vk)$ is the Fourier transform of $T^L_\parallel(\vx)$, the moments of mass-weighted velocity, defined as
\be
T^L_\parallel(\vx) = \frac{ma^{-3}}{\bar{\rho}}\int d^3\vp f(\vx,\vp)v^L_\parallel,
\label{eq:vel_moments}
\ee
where $f(\vx,\vp)$ is the phase space distribution function of particles at a phase-space position $(\vx,\vp)$.
For a single stream at ${\bf x}$, equation (\ref{eq:vel_moments}) is simply written as $T_\parallel^L(\vx)= \left[1+\delta(\vx)\right]v_\parallel^L(\vx)$ and it is obviously the $L=1$ case if equal to equation (\ref{eq:momentum}).
The equations above are applied not only to dark matter but also to biased objects \cite{Okumura:2012b}. 
The power spectrum in redshift space can be expressed as an infinite sum of cross-power spectra between mass weighted velocity moments in real space, as
\begin{eqnarray}
  P^{s}_{00}(\vk)
  &=&\sum_{L=0}^{\infty}\sum_{L'=0}^{\infty}\frac{\left(-1\right)^{L'}}{L!~L'!}
  \left(\frac{i k\mu}{\cal H}\right)^{L+L'} P_{LL'}^r(\vk)   \label{eq:p_ss} ~,
\end{eqnarray}
where $(2\pi^3)P_{LL'}^r(\vk)(\vk-\vk')=\langle T_\parallel^{L}(\vk)T_\parallel^{*L'}(\vk')\rangle$. 
Note that we explicitly put the subscript $00$ to $P^{s}$ to clarify that the so-cold redshift-space power spectrum is the auto correlation of the zeroth velocity moments in redshift space, since in the next subsection we discuss the power spectra of velocity moments in redshift space. 

By performing helicity decomposition, \cite{Seljak:2011} determined the angular dependence of 
$P_{LL'}$ and the redshift-space power spectrum was shown to be  
\bey
  P^{s}_{00}(\vk)&=&\sum_{L=0}^{\infty}\frac{1}{L!^2}\left(\frac{ k}{\cal H}\right)^{2L} \sum_{j=2L}^{4L}P^{(j)}_{LL}(\vk)\mu^{j} +
  2\sum_{L=0}^{\infty}\sum_{L'>L}\frac{\left(-1\right)^{L}}{L!~L'!}
  \left(\frac{i k}{\cal H}\right)^{L+L'} \sum_{j=(L+L') {\rm or} (L+L'+1)}^{2(L+L')}P^{(j)}_{LL'}(\vk)\mu^{j} \nonumber\\
 &\equiv& \sum_{j=0,2,4,\cdots} P_{00,\mu^j}^s(k)\mu^{j}  \label{eq:p_ss_ang} ~,
\eey
where $P^{(j)}_{LL'}$ are coefficients in expansion in powers of $\mu^j$ of contributions of $P_{LL'}^r$ to $P^s_{00}$. The coefficients $P^s_{00,\mu^j}$ are the moments expanded in terms of $\mu^2$, an alternative expansion to the commonly-used Legendre multipole expansion, 
\be
P_{00}^s(\vk)=\sum_{l=0,2,4,\cdots} P_{00,l}^s(k){\cal P}_l(\mu), \ \ \ \ \ \ P_{00,l}^s(k)=(2l+1)\int^1_0 P_{00}^s(\vk){\cal P}_l(\mu)d\mu.
\ee

We already know that the redshift-space power spectrum $P^s_{00}$ is given by the velocity statistics above as well as the density-density spectrum $P_{00}$ 
in real space under the nonlinear Kaiser 
approximation \cite{Kaiser:1987, Scoccimarro:2004, Okumura:2012b},
\be
P^{s,h}_{00}(\vk) = P^{r,h}_{00}(k) + 2f\mu^2\left(\frac{-ik}{{\cal H}\mu f}\right)P^{r,h}_{01}(\vk)+f^2\mu^4\left(\frac{k}{{\cal H}\mu f}\right)^2 P^{r,h}_{11}(\vk).  \label{eq:nl_kaiser}  
\ee 
If we additionally assume standard linear theory in equation (\ref{eq:nl_kaiser}), 
we obtain the original linear Kaiser formula, as
\be  P^{s,h}_{00,{\rm lin}}(\vk)=\left( 1+f\mu^2/b \right)^2 P_{00,{\rm lin}}^{h}(k)=\left( b+f\mu^2 \right)^2 P_{00,{\rm lin}}^{r,m}(k). \label{eq:kaiser}
\ee


\subsection{Redshift-space distortions of momentum fields}
\label{subsec:redshift_space_distortions}

A direct observable in a peculiar velocity survey are in fact peculiar velocities in redshift space, which are affected by redshift-space distortions. The position of a particle in redshift space $\vs$ is distorted by its peculiar velocity along the line of sight as $\vs=\vx+\hat{z}v_\parallel/{\cal H}$. 
Thus we can define the redshift-space momentum field as
\be
p_{\parallel}^s(\vs) = \left[ 1+\delta^s (\vs)\right] v_\parallel^s (\vs).  \label{eq:momentum_red}
\ee
By extending the phase-space distribution function approach to the redshift-space momentum field, the momentum density along the line of sight in redshift space is given by
\bey
   p^s_\pp(\VEC{s})=\frac{ma^{-3}}{\bar{\rho}}\int{d^3p~d^3x~\frac{p_\pp}{ma}f\left(\VEC{x},\VEC{p}\right)\df^D\left(\VEC{s}-\VEC{x}-\hat{z}\frac{v_\pp}{\mathcal{H}}\right)}=
                   \frac{ma^{-3}}{\bar{\rho}}\int{d^3p~\frac{p_\pp}{ma}f\left(\VEC{s}-\hat{z}\frac{v_\parallel}{\mathcal{H}},\VEC{p}\right)}. 
\label{eq:RosR}
\eey
By Fourier transforming equation (\ref{eq:RosR}), we get
\bey
    p^s_\pp (\VEC{k})&=&\frac{ma^{-3}}{\bar{\rho}}\int{d^3x~d^3p~\frac{p_\pp}{ma}f\left(\VEC{x},\VEC{p}\right)e^{i(\VEC{k}\cdot\VEC{x}+k_\pp v_\pp/\mathcal{H})}}\nonumber\\
                   &=&\frac{ma^{-3}}{\bar{\rho}}\int{d^3x~e^{i\VEC{k}\cdot\VEC{x}}}~\int{d^3p~\frac{p_\pp}{ma}f(\VEC{x},\VEC{p})e^{ik_\parallel v_\parallel/\mathcal{H}}}. 
\label{eq:RosK}
\eey
Expanding the second integral in equation (\ref{eq:RosK}) as a Taylor series in 
$k_\parallel v_\parallel/\mathcal{H}$, we get
\bey
    \frac{ma^{-3}}{\bar{\rho}}\int{d^3p~\frac{p_\pp}{ma}f\left(\VEC{x},\VEC{p}\right)}e^{ik_\pp v_\pp/\mathcal{H}} &=& \frac{ma^{-3}}{\bar{\rho}}
    \int{d^3q~f\left(\VEC{x},\VEC{p}\right)} \sum_{L=0}\frac{1}{L!}\left(ik_\pp /\mathcal{H}\right)^Lv_\pp^{L+1}\nonumber\\
    &=&\sum_{L=0}\frac{1}{L!}\left(\frac{ik_\pp}{\mathcal{H}}\right)^LT^{L+1}_\pp(\VEC{x}).
\eey
The Fourier component of the momentum  fluctuation in redshift space is
\be
    p^s_\pp(\VEC{k})=\sum_{L=0}\frac{1}{L!}\left(\frac{ik_\pp}{\mathcal{H}}\right)^LT^{L+1}_\pp(\VEC{k})
    =\sum_{L=1}\frac{1}{(L-1)!}\left(\frac{ik_\pp}{\mathcal{H}}\right)^{L-1}T^{L}_\pp(\VEC{k}).
\label{eq:psk}
\ee
This is the expression for the momentum density in redshift space corresponding to that for the mass density 
(equation \ref{eq:delta_k}).


\subsection{Momentum power spectra in redshift space} \label{subsec:redshift_power_spectrum}

As is the case with real space (equations (\ref{eq:momentum_cros_real}) and (\ref{eq:momentum_auto_real}) ), 
the redshift-space overdensity-momentum and momentum-momentum power spectrum are
defined  
as\footnote{Note that unlike the definition here, we used the definition of $P_{LL'}^s$ (or $P_{LL'}^{ss}$) to describe contributions of $P_{LL'}$ in real space to the redshift-space power spectrum $P^s(\vk)$ in our series of papers on RSD \cite{Seljak:2011, Okumura:2012, Okumura:2012b, Vlah:2012}. },
\bey
\left\langle\df^s(\VEC{k})| p^{s*}_\pp(\VEC{k}')\right\rangle=(2\pi)^3 P^{s}_{01}(\VEC{k})\df^D(\VEC{k}-\VEC{k}'),\nonumber\\ 
\left\langle p^s_\pp(\VEC{k})| p^{s*}_\pp(\VEC{k}')\right\rangle=(2\pi)^3 P^{s}_{11}(\VEC{k})\df^D(\VEC{k}-\VEC{k}').
\eey
Equations (\ref{eq:delta_k})  and (\ref{eq:psk}) give,
\bey
    P^{s}_{01}(\VEC{k})&=&\sum_{L=0}\sum_{L'=0}\frac{(-1)^{L'}}{L!L'!}\left(\frac{ik_\pp}{\mathcal{H}}\right)^{L+L'}P_{LL'+1}^r(\VEC{k})\nonumber\\
    &=&i\frac{\mathcal{H}}{k_\pp}\sum_{L=0}\sum_{L'=1}\frac{(-1)^{L'}}{L!L'!}\left(\frac{ik_\pp}{\mathcal{H}}\right)^{L+L'} L' ~ P_{LL'}^r(\VEC{k}),\nonumber\\
   P^{s}_{11}(\VEC{k})&=&\sum_{L=0}\sum_{L'=0}\frac{(-1)^{L'}}{L!L'!}\left(\frac{ik_\pp}{\mathcal{H}}\right)^{L+L'}P_{L+1L'+1}^r(\VEC{k})\nonumber\\
    &=&\left(\frac{\mathcal{H}}{k_\pp}\right)^2\sum_{L=1}\sum_{L'=1}\frac{(-1)^{L'}}{L!L'!}\left(\frac{ik_\pp}{\mathcal{H}}\right)^{L+L'} LL' ~ P_{LL'}^r(\VEC{k}).
\label{eq:Pss}
\eey
As noted in \cite{Seljak:2011} $P_{LL'}^r(\VEC{k})=P_{L'L}^r(\VEC{k})^*$. Thus only the terms $P_{LL'}^r(\VEC{k})$ with $L\leq L'$ need to be considered, 
each of which comes with a factor of 2 if $L\not=L'$ and 1 if $L=L'$. If we introduce $\mu=k_\pp/k=\text{cos}\theta$, we can write,
\bey
    -i\frac{k\mu}{\mathcal{H}}P^{s}_{01}(\VEC{k})&=&\sum_{L=1}\frac{(-1)^L}{L!}\left(\frac{ik\mu}{\mathcal{H}}\right)^L L~P_{0L}^r(\VEC{k})
                   +\sum_{L=1}\frac{1}{(L!)^2}\left(\frac{k\mu}{\mathcal{H}}\right)^{2L}L~P_{LL}^r(\VEC{k})\nonumber\\
                  &&~~~ +2Re\sum_{L=1}\sum_{L'>L}\frac{(-1)^{L'}}{L!L'!}\left(\frac{ik\mu}{\mathcal{H}}\right)^{L+L'}L'~P_{LL'}^r(\VEC{k}), \label{eq:P01ss}\\
   \left(\frac{k\mu}{\mathcal{H}}\right)^2P^{s}_{11}(\VEC{k})&=&\sum_{L=1}\frac{1}{(L!)^2}\left(\frac{k\mu}{\mathcal{H}}\right)^{2L}L^2P_{LL}^r(\VEC{k}) \nonumber \\
                  &&~~~ +2Re\sum_{L=1}\sum_{L'>L}\frac{(-1)^{L'}}{L!L'!}\left(\frac{ik\mu}{\mathcal{H}}\right)^{L+L'}LL'~P_{LL'}^r(\VEC{k}).
\label{eq:P11ss}
\eey
Considering the helicity decomposition as we did for $P^s_{00}$ in equation (\ref{eq:p_ss_ang}), we have
\be
P^s_{01}(\vk)=\sum_{j=1,3,\cdots} P_{01,\mu^j}^s(k)\mu^{j}, \ \ \ \ \ \ 
P^s_{11}(\vk)=\sum_{j=2,4,\cdots} P_{11,\mu^j}^s(k)\mu^{j}  \label{eq:p_ll_ang}.
\ee

On large scales where linear theory can be applied, peculiar velocity statistics in redshift space, $P^{s,h}_{01}$ and $P^{s,h}_{11}$, can be expressed as,
\bey
   -i\frac{k\mu}{\mathcal{H}}P^{s,h}_{01}(\VEC{k})&=&-i\frac{k\mu}{\mathcal{H}}P_{01}^{r,h}(k)+\left(\frac{k\mu}{\mathcal{H}}\right)^2P_{11}^{r,h}(k)
   =f\mu^2(b+f \mu^2)P_{00,{\rm lin}}^{r,m}(k), \label{eq:p01s_lin} \\
   \left(\frac{k\mu}{\mathcal{H}}\right)^2P^{s,h}_{11}(\VEC{k})&=&
   \left(\frac{k\mu}{\mathcal{H}}\right)^2P^{r,h}_{11}(\VEC{k})=\mu^4 f^2P_{00,{\rm lin}}^{r,m}(k). \label{eq:p11s_lin}
\eey
We thus see that even in linear theory, the density-velocity correlation is not just proportional to $\mu$, but has an additional $b+f\mu^2$ dependence that arises from RSD. 
This has usually been ignored in previous analyses \cite[e.g.,][]{Hand:2012}. 
The expression of equation (\ref{eq:p01s_lin}) is the same as that presented by \cite{Burkey:2004} who considered the cross correlation of redshift-space density and real-space velocity in the linear regime, because the velocity field in redshift space is equivalent to that in real space for the linear theory limit (see equations (\ref{eq:psk}) and (\ref{eq:p11s_lin}), and see also \cite{Koda:2013}).


\section{$N$-body simulations}\label{sec:sim}
As in our previous works \cite{Okumura:2012, Okumura:2012b}, we use a series of $N$-body simulations of the $\Lambda$CDM cosmology seeded with Gaussian initial conditions \cite{Desjacques:2009}. 
The primordial density field is generated using the matter transfer function by CAMB \cite{Lewis:2000}.
We adopt the standard $\Lambda$CDM model with $\Omega_m=1-\Omega_\Lambda=0.279$, $\Omega_b=0.0462$, 
$h = 0.7$, $n_s = 0.96$, $\sigma_8=0.807$.
We employ $1024^3$ particles of mass $m_p = 2.95\times 10^{11}h^{-1}M_\odot$ in a cubic box of side $1600\himpc$. The positions and velocities of all the dark matter particles are output at $z = 0$, 0.509, 0.989, and 2.070, which are quoted as $z = 0$, 0.5, 1, and 2 in what follows for simplicity. 
We use 12 independent realizations in order to reduce the statistical scatters. 

Dark matter halos are identified using the
friends-of-friends algorithm with a linking length equal to 0.17 times
the mean particle separation.  We use all the halos with equal to or
more than 20 particles.  In order to investigate the halo mass
dependences of the velocity statistics, each dark matter halo
catalog is divided into subsamples according to the halo mass. 
The subsamples are labelled as bin1, bin2, $\cdots$, from the least massive halos.
Since the number density of halos is
smaller at higher redshifts, we construct 4 halo subsamples at $z=0$
and 0.5, and 3 subsamples at $z=1$. We do not consider the analysis of halos at $z=2$. 
The values of the linear bias parameter $b_1$ determined using $\chi^2$ statistics for $b(k)=P_{00}^{r,mh}(k)/P_{00}^{r,mm}(k)$ at $0.01\leq k \leq 0.04\hmpci$ are 
$b_1=1.17$, 1.45, 2.02 and 3.03 at $z=0$, 
$b_1=1.64$, 2.16, 3.11 and 4.85 at $z=0.5$, and 
$b_1=2.31$, 3.17 and 4.64 at $z=1$. 
$P_{00}^{mh}$ is the cross power spectrum of matter and halo density fields. 
See table 1 of \cite{Okumura:2012b} for the detail of the halo catalogs. 

We measure the power spectra of velocity moments for dark matter and halos from our simulation samples following \cite{Okumura:2012, Okumura:2012b}.  
We assign the number-weighted velocity moments in real space and in redshift space
on $1024^3$ grids using a cloud-in-cell interpolation method according
to the positions of particles. In measuring the
velocity field in redshift space, we distort the positions of particles
along the line-of-sight according to their peculiar velocities before
we assign them to the grid.  We regard each direction along the three
axes of simulation boxes as the line of sight and the statistics are
averaged over the three projections of all realizations for a total of
36 samples. We use a fast Fourier transform to measure the Fourier
modes of the number-weighted
velocity moment fields in real space $T_\parallel^{L}(\vk)$ and 
in redshift space $T_\parallel^{s,L}(\vk)$.  Then
the power spectrum $P_{LL'}^{r}(\vk)$ and $P_{LL'}^{s}(\vk)$ 
are measured by multiplying the modes of the two
fields (or squaring in case of auto-correlation) and averaging over
the modes within a bin.  
The momentum field is normalized by ${\cal H}(a)=aH(a)$, thus the momentum power spectra $P_{01}$ and $P_{11}$ respectively have the unit of $[\himpc ]^4$ and $[\himpc ]^5$. 
To show the error of the mean for measured statistics, the scatter among realizations needs to be divided by the square root of the number of the realizations. Three measurements along different line-of-sight in each of 12 realizations are, however, not fully independent. To be conservative, we thus present the measured dispersion divided by $\sqrt{12}$ as the error of the mean in the following analysis. 


\section{Fourier-space analysis}\label{sec:analysis}

The purpose of this paper is to present the momentum cross and auto power spectrum 
for dark matter and halos in redshift space measured from $N$-body simulations and predict them 
using linear theory and one-loop nonlinear perturbation theory. 
As we did in \cite{Vlah:2012, Vlah:2013}, we consider terms that contribute to the redshift-space velocity statistics, $P_{01}^s$ and $P_{11}^s$, up to one loop in Eulerian PT, so we have $P^r_{LL'}$ up to $L+L'\leq 4$ and
\bey
-i\frac{k\mu}{\mathcal{H}}P^{s}_{01}(\VEC{k})&=&
-i\frac{k\mu}{\mathcal{H}}P^r_{01}(\vk)
+\left(\frac{k\mu}{\mathcal{H}}\right)^2P^r_{11}(\vk)
-\left(\frac{k\mu}{\mathcal{H}}\right)^2P^r_{02}(\vk)
+\frac{i}{2}\left(\frac{k\mu}{\mathcal{H}}\right)^3P^r_{03}(\vk) \nonumber \\
&&-2i\left(\frac{k\mu}{\mathcal{H}}\right)^3P^r_{12}(\vk)
+\frac{1}{6}\left(\frac{k\mu}{\mathcal{H}}\right)^4P^r_{04}(\vk)
-\left(\frac{k\mu}{\mathcal{H}}\right)^4P^r_{13}(\vk)
+\frac{1}{2}\left(\frac{k\mu}{\mathcal{H}}\right)^4P^r_{22}(\vk) , \label{eq:p01s_pt} \\
   \left(\frac{k\mu}{\mathcal{H}}\right)^2P^{s}_{11}(\VEC{k})&=&
   \left(\frac{k\mu}{\mathcal{H}}\right)^2P^r_{11}(\VEC{k})
-2i\left(\frac{k\mu}{\mathcal{H}}\right)^3P^r_{12}(\vk)
-\left(\frac{k\mu}{\mathcal{H}}\right)^4P^r_{13}(\vk)
+\left(\frac{k\mu}{\mathcal{H}}\right)^4P^r_{22}(\vk). \label{eq:p11s_pt}
\eey
Among these terms in equation (\ref{eq:p01s_pt}), all the terms that contribute to $-i(k\mu/{\cal H})P^s_{01}$ as $\mu^2$ and $\mu^4$, except for $P^r_{04}$, can be formulated using one-loop PT for dark matter \cite{Vlah:2012} and halos \cite{Vlah:2013}. 
The $\mu^4$ term in $(k\mu/{\cal H})^4P_{04}^r$ is of two-loop order, but we take into account the PT modeling of it because, as shown in \cite{Vlah:2012}, it has a non-negligible contribution. 
We thus include all the terms that have the angular dependence of $\mu$ and $\mu^3$ to $P^s_{01}$ in the following analysis, i.e., $\mu^2$ and $\mu^4$ to $-i(k\mu/{\cal H})P^s_{01}$.
Likewise, we take into account all the terms that contribute to $\mu^0$ and $\mu^2$ in $P^s_{11}$. 
All the linear theory predictions presented below are computed using CAMB \cite{Lewis:2000}.

As discussed in \cite{Vlah:2012}, modeling the matter power spectrum in real space, $P_{00}^{r,m}$, 
plays an important role to predict both $P_{LL'}^{r,m}$ and $P_{LL'}^{r,h}$. 
The matter power spectrum in real space is the most fundamental spectrum and 
there are many attempts to model it based on perturbation theory for quasi-nonlinear scales 
\cite[e.g.,][]{Bernardeau:2002, Crocce:2006, McDonald:2007, Taruya:2008, Pietroni:2008}
and on calibration using $N$-body simulations for fully nonlinear scales \cite[e.g.,][]{Smith:2003, Heitmann:2010, Takahashi:2012}. 
In this paper, rather than modeling it we assume that we have a perfect theoretical model for $P_{00}^{r,m}$, namely, 
the modeled $P_{00}^{r,m}$ is equal to that measured from the $N$-body simulation.
With this assumption we also have perfect model for $P_{01}^{r,m}$ because the scalar part of the momentum field is related to the time derivative of the 
matter density field (see equation (3.7) of \cite{Seljak:2011}). 
Note that we still use nonlinear PT for the vector part of $P_{11}^{r,m}$ that was shown to be in a good agreement with $N$-body result \cite{Vlah:2012}. 
In the following analysis, thus, we use values of $P_{00}^{r,m}$, $P_{01}^{r,m}$ and the scalar part of $P_{11}^{r,m}$ measured 
from $N$-body simulations as our prediction and use nonlinear PT at one-loop order for the other $P_{LL'}^{r,m}$'s and at two-loop for $P_{04}^{r,m}$. In this way we can investigate the accuracy of nonlinear PT on prediction of nonlinear RSD, since all the possible deviations from the $N$-body measurement come from the imperfect modeling of nonlinear RSD, as well as the imperfect modeling 
of nonlinear bias in the case of analysis of dark matter halos.

As we have seen in \cite{Okumura:2012b, Vlah:2012}, all the $P_{LL'}^r$'s in equations (\ref{eq:p01s_pt}) and (\ref{eq:p11s_pt}), except $P_{01}^r$ and $P_{11}^r$, have terms proportional to velocity dispersion $\sigma_v^2$. Briefly, at one loop order, the isotropic part of $P_{02}^r$ contains a term proportional to $\sigma_v^2 P_{00}^r$, the correlators $P_{12}^r$ and $P_{03}^r$ contain a term proportional to $\sigma_v^2 P_{01}^r$, $P_{13}^r$ is simply proportional to $\sigma_v^2 P_{01}^r$, and $P_{22}^r$ contains a term proportional to $\sigma_v^2 P_{02}^r$ and another proportional to $\sigma_v^4 P_{00}^r$. The correlator $P_{04}^r$ contains a term proportional to $\sigma_v^2 P_{02}^r$ and another proportional to $\sigma_v^4 P_{00}^r$ at two loop order. Using perturbation theory, the velocity dispersion is evaluated as 
\be
\sigma_v^2=\frac{1}{3}\int\frac{d^3q}{(2\pi)^3}\frac{P_{\theta\theta}(q)}{q^2},
\ee
where $P_{\theta\theta}$ is the velocity divergence power spectrum. 
While linear theory gives $\sigma_v\simeq 6fD\himpc$, which is a good approximation for halo velocities \cite{Vlah:2013}, 
this does not properly take account of small scale contributions which come from within virialized halos 
where PT cannot be applied. 
In order to take into account the nonlinear contributions to the velocity dispersion, we add the corrections
$\sigma_{LL'}^2$ to $\sigma_v^2$ in $P_{LL'}^{r,m}$ as $\sigma_v^2 \rightarrow \sigma_v^2+\sigma_{LL'}^2$.
Rather than treating the nonlinear correction $\sigma_{LL'}$ as free parameter, 
we simply adopt the best fit parameters of $\sigma_{LL'}$, which were obtained for the real-space 
$P^{r,m}_{LL'}$ and shown in Table 1 of \cite{Vlah:2012}. 


\subsection{Dark matter in real space and in redshift space}
\label{sec:analysis_dm}

In figure \ref{fig:pkmu_01_11} we show the density-momentum cross power spectra and the momentum 
auto power spectra measured for dark matter and halos
as functions of $(k,\mu)$ both in real space and redshift space at $z=0$. 
We adopt the constant $\mu$ binning into five bins between $0\leq \mu \leq 1$, but only four $\mu$ bins among 
the five are plotted, $0\leq \mu \leq 0.2$ (green), $0.2\leq \mu \leq 0.4$ (yellow), $0.4\leq \mu \leq 0.6$ (red) 
and $0.8\leq \mu \leq 1$ (blue). 
The corresponding linear theory predictions are shown as the dotted line with the same color. 
Our nonlinear PT predictions are shown as the solid lines. 
In the rest of this section we will see the results in detail. 
First let us focus on the results for dark matter that are shown in the left column of figure \ref{fig:pkmu_01_11}, while
those for halos will be presented in the next subsection. 

\subsubsection{Cross power spectrum between density and momentum}
\label{sec:analysis_dm_p01}
\begin{figure}[bt] 
  \centering
  \includegraphics[width=.985\linewidth]{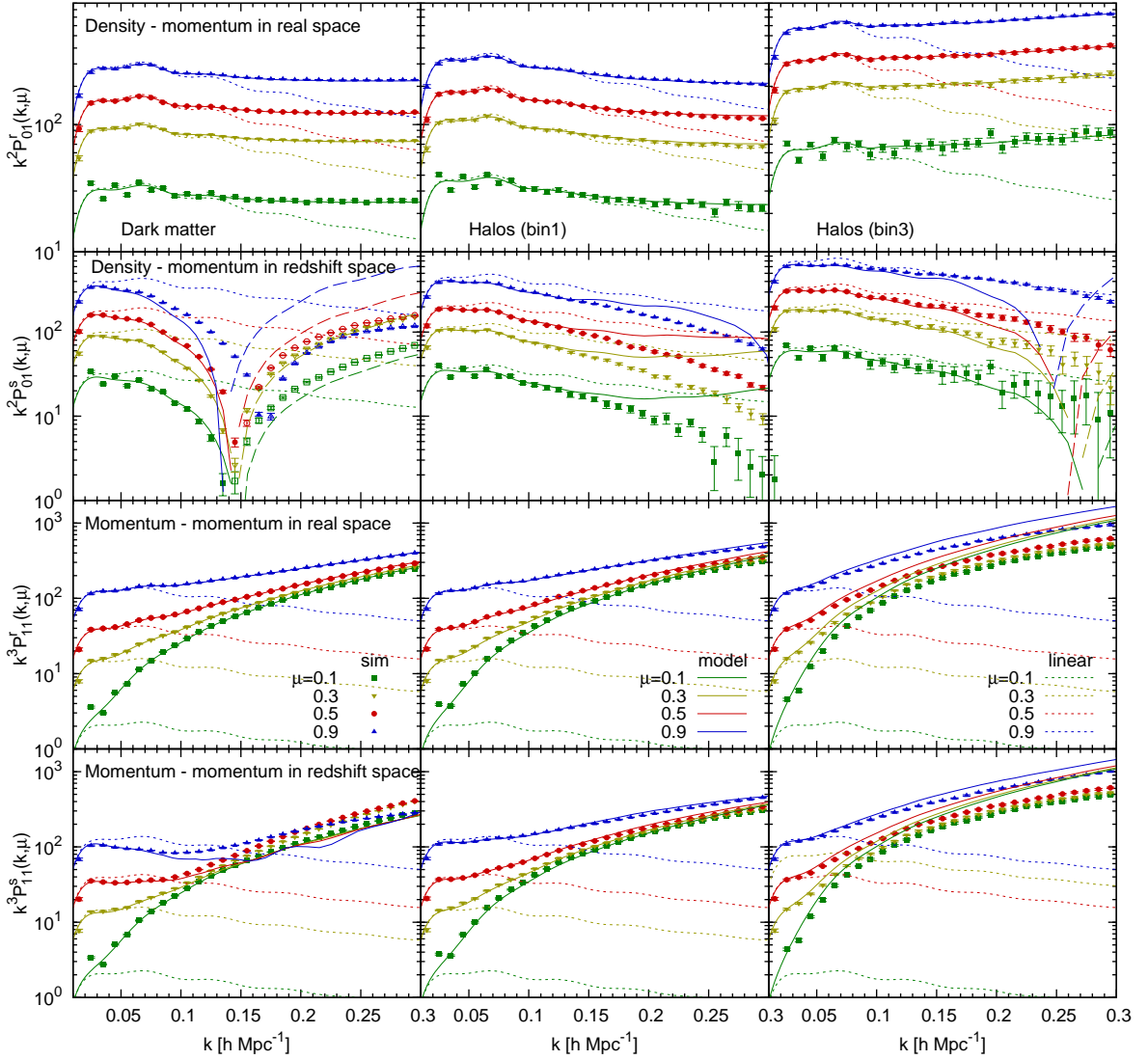}
\caption{Density-momentum cross power spectra in real
  space $P_{01}(k,\mu)$ ({\it first row}) and in redshift
  space $P_{01}^{s}(k,\mu)$ ({\it second row}), and 
  momentum auto power spectra in real space $P_{11}(k,\mu)$ ({\it third row}) and in redshift
  space $P_{11}^{s}(k,\mu)$ ({\it bottom row}), all measured at $z=0$.
 From left to right we show the results for dark matter, halos with mass bin 1, and halos with mass bin 3. 
  $P_{01}$ and $P_{11}$ are respectively multiplied by $k^2$ and $k^3$ for clarity. 
  The points with different color show the results with different $\mu$ values, quoted in the third row, with the width of $\pm 0.1$.
The sold lines show our PT predictions and the negative values are shown as the dashed lines. 
The corresponding linear theory predictions 
are shown as the dotted line with the same color.}
\label{fig:pkmu_01_11}
\end{figure}

At the top of the left column of figure \ref{fig:pkmu_01_11} we show the density-momentum power spectrum in real space $P^{r,m}_{01}(k,\mu)$, multiplied by $k^2$ for dark matter.  
This is the quantity that has been already computed and presented in \cite{Okumura:2012, Vlah:2012}, and we show it here to compare with the corresponding redshift-space spectrum.
As we have seen in \cite{Okumura:2012, Vlah:2012}, this statistics approaches linear theory at large scales shown as the dotted lines, while for $k>0.1\hmpci$ it exceeds linear theory. 
Because the $N$-body measurement of $P_{01}^{r,m}$ is used as our theoretical prediction, the prediction shown as the solid lines coincides with the measured $P_{01}^{r,m}$ by construction. 

\begin{figure}[bt]
\centering
\includegraphics[width=.985\textwidth]{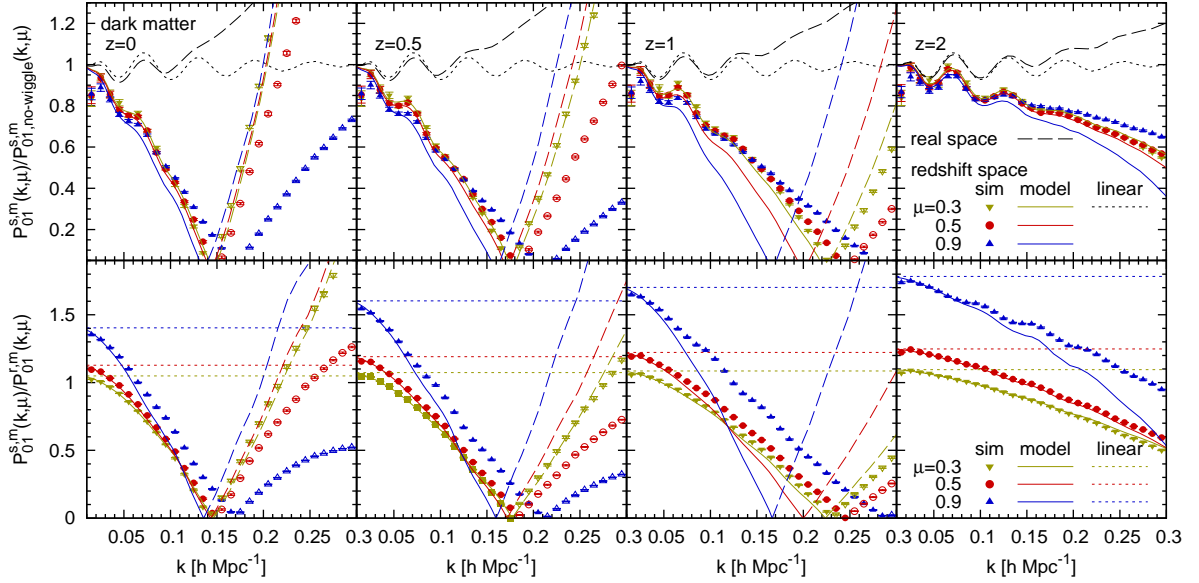}
\caption{{\it Top row}: Density-momentum cross power spectra for dark matter in redshift space divided by the no-wiggle spectra \cite{Eisenstein:1998}, $P_{01}^{s,m}/P_{01,\text{no-wiggle}}^{s,m}$. 
From the left to right, results are shown for $z=0$, 0.5, 1 and 2. 
The solid lines show our PT predictions and the negative values are shown as the dashed lines. 
For comparison, the results in real space, $P_{01}^{r,m}/P_{01,\text{no-wiggle}}^{r,m}$, are shown as the black dashed line (it is angle independent). The corresponding linear theory prediction is isotropic and equivalent in real space and redshift space, $P_{01,{\rm lin}}^{r,m}/P_{01,\text{no-wiggle}}^{r,m}=P_{01,{\rm lin}}^{s,m}/P_{01,\text{no-wiggle}}^{s,m}$, shown as the dotted lines. 
{\it Bottom row}: Ratios of the spectra in redshift space and real space, $P_{01}^{s,m}(k,\mu)/P_{01}^{r,m}(k,\mu)$. 
The corresponding nonlinear PT prediction is shown as the solid (positive) and dashed (negative) lines with the same color as the measurement, 
while linear theory prediction is as the dotted lines.}
\label{fig:pkmu_01_dm}
\end{figure}

Next, at the left panel of the second row of figure \ref{fig:pkmu_01_11} the density-momentum power spectrum for dark matter measured in redshift space, $P^{s,m}_{01}(k,\mu)$, is shown. 
There are two terms, $P^{r,m}_{01}$ and $P^{r,m}_{11}$, that have linear-order contributions to $P^{s,m}_{01}$. 
As expected, $P^{s,m}_{01}$ starts to deviate from linear theory at larger scales than $P^{r,m}_{01}$.
At $k\simeq 0.15\hmpci$ the sign of $P^{s,m}_{01}$ is changed, because of the FoG effect, mainly from $P_{02}$. 
The more detailed comparison is presented for four redshifts at the top row of figure \ref{fig:pkmu_01_dm} where we plot the ratio of $P^{s,m}_{01}$ to the corresponding linear power spectrum without BAO wiggles $P^{s,m}_{01,\text{no-wiggle}}=(1+f\mu^2)P^{r,m}_{01,\text{no-wiggle}}$. The linear theory no-wiggle power spectrum in real space $P^{r,m}_{01,\text{no-wiggle}}$ is computed using the formula of \cite{Eisenstein:1998}.
Here $P^{s,m}_{01,\text{no-wiggle}}$ has two terms, one proportional to $\mu$ and another proportional to $\mu^3$, thus the ratio $P^{s,m}_{01}/P^{s,m}_{01,\text{no-wiggle}}$ has angular dependence. One can see that except at very large scales $k\sim 0.01 \hmpci$, linear theory cannot explain the $N$-body results at all. 
On the other hand, our nonlinear PT prediction agrees well with the $N$-body results. 
Our PT calculation also well explains the trend of the change of signs of $P^{s,m}_{01}$. 
However, at higher $\mu$, the deviation of the PT prediction from the $N$-body result starts at larger scales: 
at $\mu=0.9$ $k\sim 0.05\hmpci$ for $z=0$ and $k\sim 0.08\hmpci$ for $z=2$. 
This is expected because $-(ik\mu/{\cal H})P_{01}^s$ has the angular dependence of $\mu^{2j} \ (j=1,2,\cdots)$,
while we used terms up to $\mu^4$ using PT, and the neglected higher order terms are more significant 
for higher $\mu$. 
On the other hand, such higher-order terms are significantly suppressed for smaller $\mu$, thus 
our one-loop PT agrees with $N$-body results of $P_{01}^{s,m}$ at $\mu=0.3$ up to much smaller scales. 
For comparison, the density-momentum spectrum in real space $P^{r,m}_{01}(k,\mu)/P^{r,m}_{01,\text{no-wiggle}}(k,\mu)$ is shown in the top row of figure \ref{fig:pkmu_01_dm} as the black dashed line (angle independent).  
Since $P_{01}^r$ has angular dependence of $P_{01}^r(\vk)=\mu P_{01,\mu^1}^r(k) $, the ratio of the measured $P^{r,m}_{01}$ to the corresponding no-wiggle linear spectrum becomes isotropic. 
The ratios of the cross power spectrum in redshift space to that in real space, 
$P^{s,m}_{01}/P^{r,m}_{01}$, are shown in the bottom panels of figure \ref{fig:pkmu_01_dm}.
At the linear theory limit this ratio gives $(1+f\mu^2)$, shown as the dotted horizontal lines. 
Nonlinear PT can predict the numerical results for $\mu=0.9$ up to $k\simeq 0.05 \hmpci$ at $z=0$ and 
to $k\simeq 0.08\hmpci$ at $z=2$, and to the smallest scales probed here for $\mu=0.3$. Note that 
we use $\mu=0$ simulation results as the starting point. 

\subsubsection{Auto power spectrum of momentum}
\label{sec:analysis_dm_p11}
\begin{figure}[bt]
\includegraphics[width=.985\textwidth]{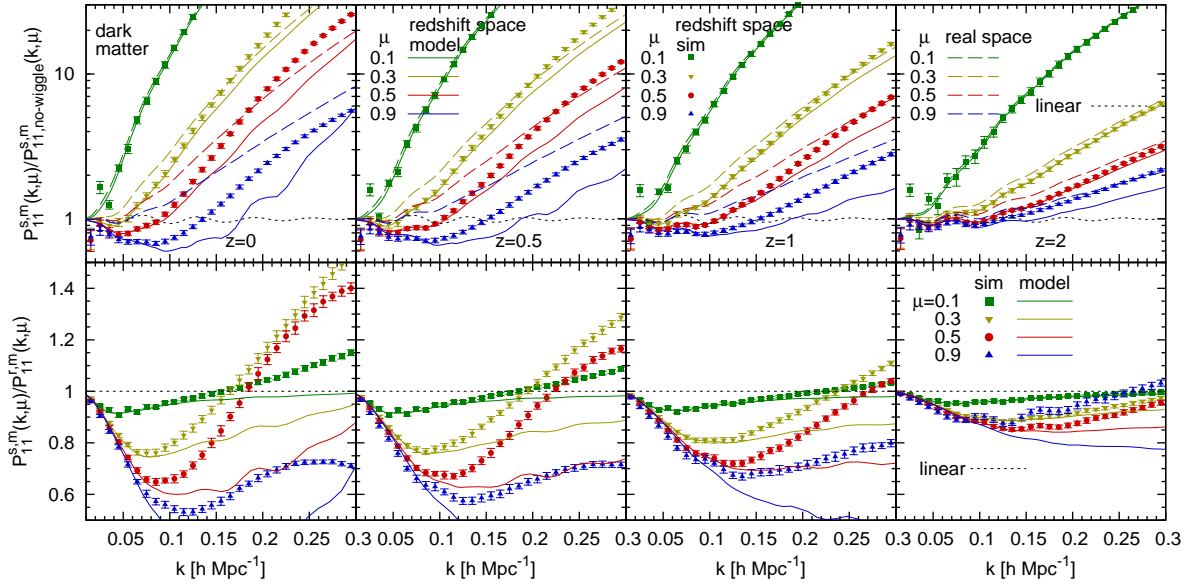}
\caption{{\it top row}: Momentum auto power spectra for dark matter in redshift space divided by the no-wiggle spectra $P_{11}^{s,m}/P_{11,\text{no-wiggle}}^{s,m}$. From the left results are shown for $z=0$, 0.5, 1 and 2. The solid lines show nonlinear PT predictions with the same color as the measurement, while  the linear theory prediction is shown as the black dotted line. For comparison the real-space result $P_{11}^{r,m}/P_{11,\text{no-wiggle}}^{r,m}$ is shown as the dashed lines. 
{\it bottom row}: Ratios of the spectra in redshift space and real space, $P_{11}^{s,m}(k,\mu)/P_{11}^{r,m}(k,\mu)$. The solid lines and the horizontal dotted lines are the nonlinear and linear theory predictions, respectively. 
}
\label{fig:pkmu_11_dm}
\end{figure}

Third, we show at the left panels of the third and fourth rows in figure \ref{fig:pkmu_01_11} the momentum auto power spectrum in real space $P^{r,m}_{11}$ and in redshift space $P^{s,m}_{11}$, respectively. 
In linear theory the two statistics are equivalent (equation (\ref{eq:p11s_lin})) and the behaviors of these measurements are indeed similar. 
As we have seen in \cite{Okumura:2012, Vlah:2012}, $P^{r,m}_{11}$ exceeds linear theory at small scales just like $P^{r,m}_{01}$ but the nonlinear effect is stronger. 
For prediction of the auto momentum spectrum in real space, we use the measurement from $N$-body simulation for the scalar part and nonlinear PT for the vector part.
Because the vector part of $P^{r,m}_{11}$ is well modeled using one-loop nonlinear PT \cite{Vlah:2012}, the measured $P^{r,m}_{11}$ perfectly agrees with our prediction. 
Unlike $P^{s,m}_{01}$, $P^{s,m}_{11}$ does not change the sign at all scales because there is no contribution from $P_{02}^{r,m}$ that has a large FoG effect. 
The momentum auto power spectra of dark matter in real space and in redshift space divided by their corresponding no-wiggle spectra are shown at the first row in figure \ref{fig:pkmu_11_dm}.
Interestingly, the momentum auto power spectrum in redshift space $P^{s,m}_{11}$ is better predicted using our nonlinear PT 
than the density-momentum cross spectrum in redshift space $P^{s,m}_{01}$. 
As is the case with $P^{s,m}_{01}$, however, the deviation of our PT prediction from the simulation result is larger at higher $\mu$, since we use only the terms up to $\mu^4$ for $(k\mu/{\cal H})^2P_{11}^s$ while 
it has the infinite angular dependence of $\mu^{2j} \ (j=1,2,\cdots)$.
The bottom row of figure \ref{fig:pkmu_11_dm} shows the ratio of the momentum auto spectra in 
redshift space and in real space, $P^{s,m}_{11}/P^{r,m}_{11}$, that is just unity in linear theory. 
Although the effect of RSD on $P_{11}$ is a fully nonlinear effect, that changes the amplitude of $P_{11}$ by 
$\sim 50\%$ at $k\simeq 0.1\hmpci$ at $z=0$ and $\sim 15\%$ even at $z=2$. 
These trends are well captured by our one-loop PT calculations. 

\begin{figure}[bt]
\includegraphics[width=.985\textwidth]{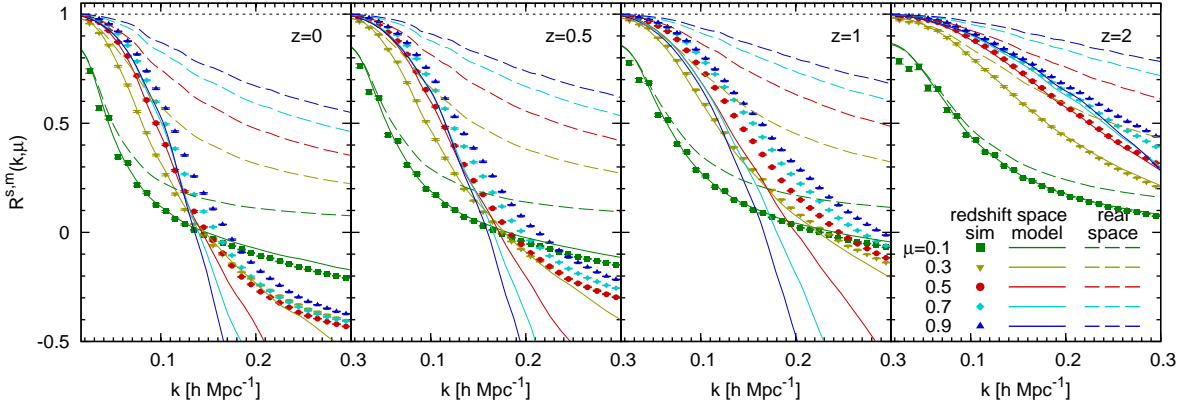}
\caption{Cross-correlation coefficient of the density and momentum for dark matter at four redshifts 
in redshift space $R^{s,m}(k,\mu)$.  
The sold lines show our PT predictions. In linear theory the cross-correlation coefficients are simply unity,
as denoted by the black dotted lines.
For comparison the coefficient in real space is shown as the dashed lines. 
}
\label{fig:rkmu_dm}
\end{figure}

Let us consider the cross-correlation coefficient between density and momentum in redshift space, defined as,
\be 
R^{s,m} (\vk)= \frac{P^{s,m}_{01}(\vk)}{\sqrt{P^{s,m}_{00}(\vk)P^{s,m}_{11}(\vk)}}.
\ee 
In linear theory the dark matter density field is perfectly correlated with the momentum field thus 
the cross-correlation coefficient is just unity.
We show in figure \ref{fig:rkmu_dm} the cross-correlation coefficient for dark matter in redshift space as functions of $(k,\mu)$.
For comparison we also show the cross-correlation coefficient defined in real space, $R^{r,m}$, that is a similar quantity presented by \cite{Carlson:2009} for the velocity field instead of the momentum field.  
Because the $N$-body measurement and prediction of $R^{r,m}$ are almost identical, we show only the latter as the dashed lines. 
The angular dependence of $R^{r,m}$ comes from the nonlinear, isotropic part of $P^{r,m}_{11}$.
If only the scalar part of $P^{r,m}_{11}$ is used, the coefficient becomes isotropic $R^{r,m}(\vk)=R^{r,m}(k)$ \cite{Okumura:2012b}.
Because of the decoherence of density and momentum fields, $R^{r,m}$ approaches 0 at small scales. 
As for the cross-correlation coefficient in redshift space, $R^{s,m}(k,\mu)$, since the momentum field here is correlated with the redshift-space density field that is distorted by the momentum field, it does not approach 0 but goes to negative values. 
The trend is well captured by our nonlinear PT prediction. 
However, at low redshifts the deviation from $N$-body results gets larger and larger for higher $k$. 

\subsection{Halos in redshift space}\label{sec:result_halos}
\label{sec:analysis_halo}
Let us move onto results obtained for biased objects, dark matter halos. 
As mentioned above, halo velocity dispersion is well predicted by linear theory with weak dependence of halo mass, so  we use the linear theory prediction, $\sigma_v/fD= 6\himpc$.
There is an additional issue to model halo $P_{LL'}^{h}$, that is halo biasing. 
Following \cite{Vlah:2013} we adopt an Eulerian biasing model \cite{McDonald:2009a}
\be
\delta_h[\delta_m]=b_1\delta_m(\vx)+\frac{b_2}{2}\left[\delta_m^2(\vx)-\langle\delta_m^2\rangle\right]
+\frac{b_s}{2}\left[s(\vx)^2-\langle s^2\rangle\right] +\frac{b_3}{6}\delta_m^3(\vx), \label{eq:bias_model}
\ee
where $b_s$ is the nonlocal bias term that comes from the tidal tensor.
With bias renormalization $b_3$ can be absorbed into $b_1$, and $b_s$ can be related to $b_1$ through $b_s=-\frac{2}{7}(b_1-1)$ \cite{Baldauf:2012}. 
Thus we will consider a set of the two bias parameters $(b_1, b_2)$ for each halo subsample. 
We use the models of $P_{LL'}^{h}$ with the best fit values of the halo biasing $(b_1, b_2)$ 
found by \cite{Vlah:2013}. 
As is the case in section \ref{sec:analysis_dm}, we assume that we can model the dark matter power spectrum in real space $P_{00}^m$, thus we use $P_{00}^{r,m}$, $P_{01}^{r,m}$ and the scalar part of $P_{11}^{r,m}$ measured from simulations. 

\subsubsection{Cross power spectrum between density and momentum}
\label{sec:analysis_halo_p01}
At the first row in figure \ref{fig:pkmu_01_11}, the middle and right panels show the density-momentum cross power spectra of the lightest halos (bin1) and the second most massive halos (bin3), respectively, in real space $P^{r,h}_{01}$ measured at $z=0$. 
Just like the case for dark matter, $P^{r,h}_{01}$ for halos exceeds linear theory at small scales. 
Because we studied the density-momentum power spectrum of halos in real space in detail in \cite{Okumura:2012b, Vlah:2013}, the result is shown in appendix \ref{sec:real_space} only for $z=0$ as a comparison to the corresponding redshift-space results shown below.

\begin{figure}[bt]
\includegraphics[width=.985\textwidth]{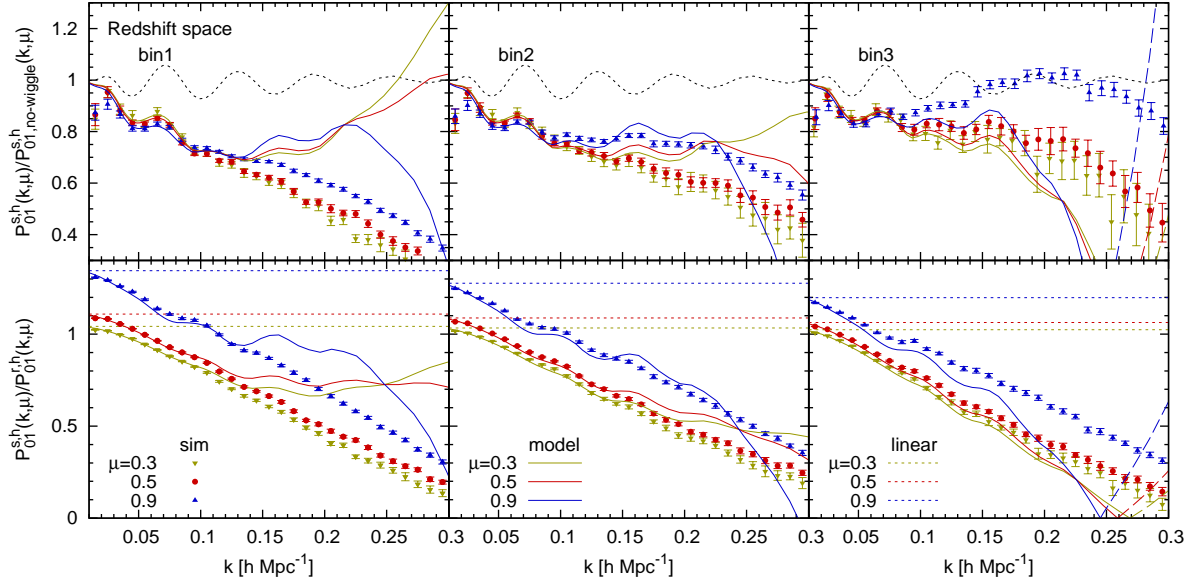}
\caption{
{\it Top row}: Density-momentum cross power spectra for halos in redshift space divided by the no-wiggle spectra at $z=0$, $P_{01}^{s,m}/P_{01,\text{no-wiggle}}^{s,m}$. {\it Bottom row}: Ratios of the spectra in redshift space and real space, $P_{01}^{s,m}(k,\mu)/P_{01}^{r,m}(k,\mu)$.
The corresponding linear theory predictions are shown as the dotted line with the same color, while 
our nonlinear PT predictions are as the solid and dashed lines for positive and negative values, respectively.
}
\label{fig:pkmu_01_z007_3}
\end{figure}

\begin{figure}[bt]
\includegraphics[width=.985\textwidth]{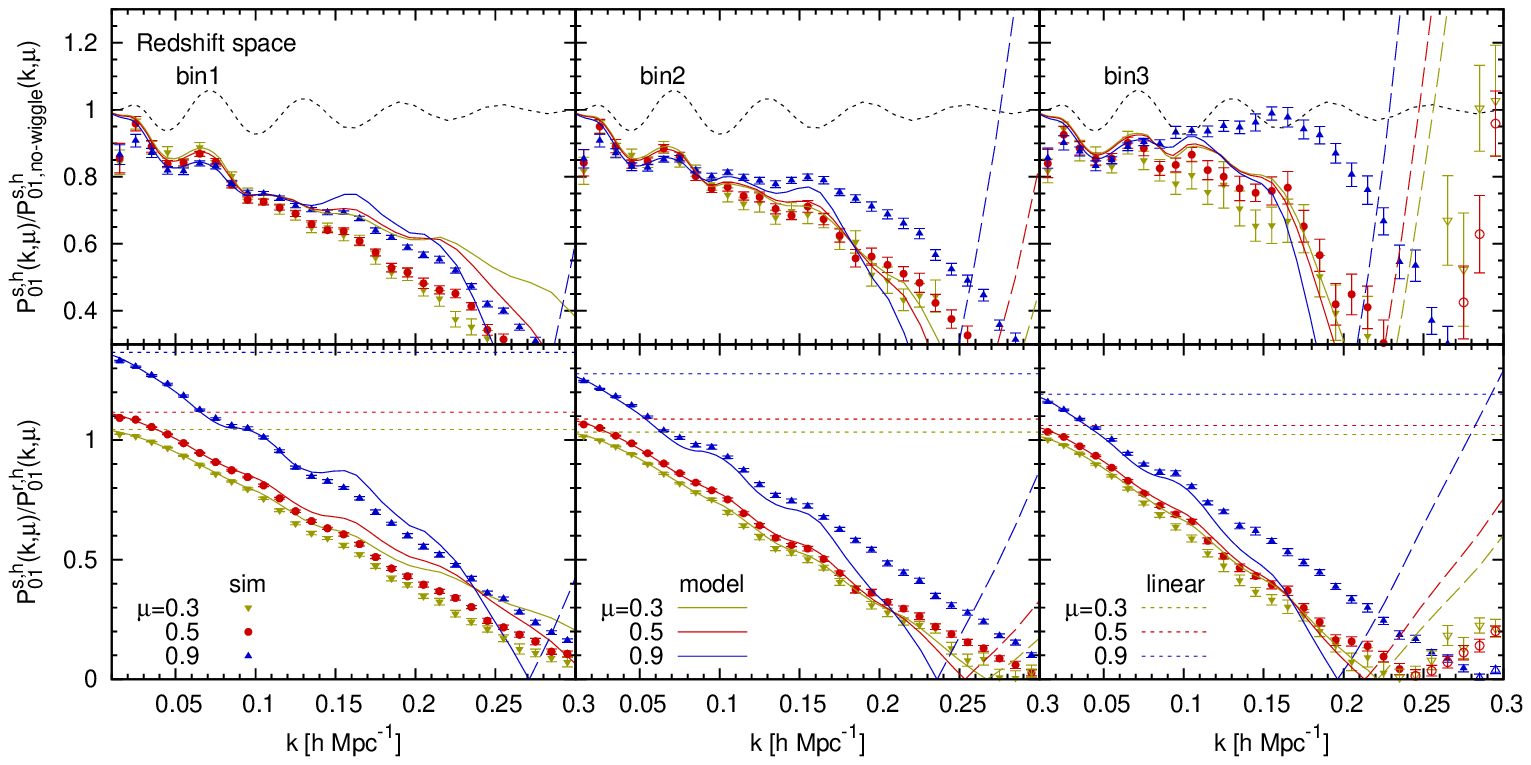}
\caption{Same as figure \ref{fig:pkmu_01_z007_3} but for $z=0.5$.
}
\label{fig:pkmu_01_z005_3}
\end{figure}

\begin{figure}[bt]
\includegraphics[width=.985\textwidth]{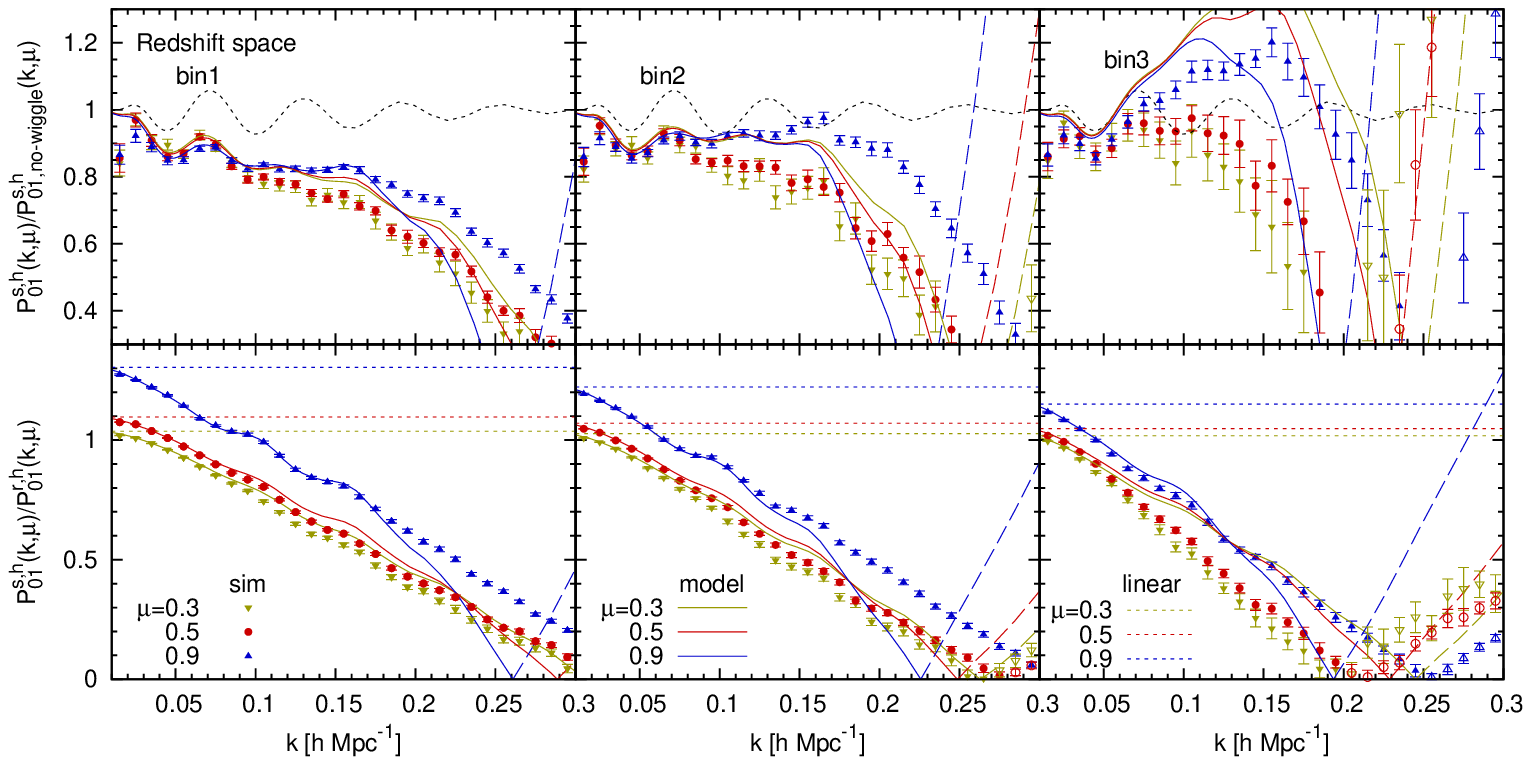}
\caption{Same as figure \ref{fig:pkmu_01_z007_3} but for $z=1$. 
}
\label{fig:pkmu_01_z004_3}
\end{figure}

The cross power spectrum for halos in redshift space $P^{s,h}_{01}(k,\mu)$ is shown at the middle and right panels of the second row in figure \ref{fig:pkmu_01_11}. The spectrum for halos is closer to linear theory than that for dark matter. It is because the power spectrum in redshift space is expanded in terms of $k\mu v_\pp/{\cal H}$, where radial velocity $v_\pp$ is much smaller for halos than for dark matter and the suppression of the amplitude by the nonlinear velocity dispersion is partially cancelled out with the nonlinear clustering as we have seen in RSD of halos \cite{Okumura:2012b}. Because each term of the expansion of $P_{01}^s$ has a different factor, e.g., $P_{11}^r+P_{02}^r$ in $P_{01}^s$ is twice as large relative to $P_{01}^r$ as that in $P_{00}^s$, the halo cross spectrum in redshift space $P_{01}^{s,h}$ changes sign at smaller scales than $P_{00}^{s,h}$. 
The halo mass dependence of $P^{s,h}_{01}$ is shown at the first rows of figures \ref{fig:pkmu_01_z007_3}, \ref{fig:pkmu_01_z005_3} and \ref{fig:pkmu_01_z004_3} for $z=0$, 0.5 and 1, respectively. As is the case with dark matter $P^{s,m}_{01}$, $N$-body results for halos disagree with linear theory prediction except for the largest scale $k\leq 0.02\hmpci$.  Although the velocity dispersion of halos is so small that linear theory can be applied, the amplitude of $P^{s,h}_{01}$ is suppressed relative to linear theory prediction for all the halo subsamples 
for all the redshifts. 
Our nonlinear prediction at one-loop level with a set of bias $(b_1,b_2)$ significantly improves the accuracy. 
Except the lowest mass bin at $z=0$, the nonlinear PT well predicts the scale dependence up to small scales 
including the scale where the sign of $P_{01}^{s,h}$ turns negative. 
The ratio of the cross power spectrum in redshift space and in real space, $P^{s,h}_{01}/P^{r,h}_{01}$, is 
shown at the bottom rows in figures \ref{fig:pkmu_01_z007_3}, \ref{fig:pkmu_01_z005_3} and 
\ref{fig:pkmu_01_z004_3}. 
The more massive halos we focus on, the more isotropic 
the ratio $P^{s,h}_{01}/P^{h}_{01}$ on linear scales becomes since the anisotropy induced by RSD 
is parameterized by $f/b$.
Although the overall trend of each simulation result is well captured by nonlinear PT, 
a careful look at the comparison reveals small differences between $N$-body results and the PT prediction. 
For example at $z=0.5$, they agree with each other for mass bin1 up to $k\sim 0.15\hmpci$ for $\mu=0.3$
and up to $k\sim 0.13\hmpci$ for $\mu=0.9$, and for mass bin3 up to $k\sim 0.2\hmpci$ for $\mu=0.3$
and up to $k\sim 0.11\hmpci$ for $\mu=0.9$. The prediction of $P_{01}^s(k,\mu)$ along the line of sight 
is less accurate than that perpendicular to the line of sight because we consider PT at one-loop level thus 
the angular dependence is truncated at $\mu^3$. 


\subsubsection{Auto power spectrum of momentum}
\label{sec:analysis_halo_p11}
Next we consider the auto-power spectrum of momentum for halos $P^{r,h}_{11}$ and $P^{s,h}_{11}$. 
They have an isotropic term thus affected by shot noise \cite{Seljak:2011, Okumura:2012b}.
In our previous works it was not important because the shot noise on $P^{r,h}_{11}$ is canceled out by that coming from the density-energy density spectrum $P^{r,h}_{02}$ when their contributions to the redshift-space power spectrum $P^{s,h}_{00}$ is considered. 
Here, because we consider the momentum spectrum itself, the shot noise needs to be estimated and subtracted. 
\cite{Seljak:2009a} defined the scale-dependent shot noise for the halo density field $\sigma_{n,00}(k)$ as 
$\delta_h(k) = b\delta_m(k)+ \sigma_{n,00}(k)$,
then the shot noise for the halo density spectrum was estimated as
\be
\sigma_{n,00}^2(k)=P_{00}^{r,h}(k)-2b_1P_{00}^{r,mh}(k)+b_1^2P_{00}^{r,m}(k),\label{eq:shot_00}
\ee
where $P_{00}^{r,mh}$ is the cross power spectrum of matter and halo density fields in real space. 
If one assumes the Poisson model the contribution of the shot noise to the halo power spectrum is simply $\sigma_{n,00}^2=1/n$, where $n$ is the halo number density.
With the analogy to the method of \cite{Seljak:2009a} above, we can estimate the shot noise for the momentum spectrum as
\be
\sigma_{n,11}^2(k)=P_{11}^{r,h}(\vk)-2P_{11}^{r,mh}(\vk)+P_{11}^{r,m}(\vk), \label{eq:shot_11}
\ee
where $P^{r,mh}_{11}$ is the real-space cross power spectrum between matter and halo momentum fields.
Here we assumed that the momentum bias is unity at the large scale limit, which is a reasonable assumption 
as is clear from \cite{Okumura:2012b}. Again, under the assumption of the Poisson model we have 
$\sigma_{n,11}^2=\langle v_{\parallel}^2 \rangle/n = \langle v_{\parallel}^2 \rangle\sigma_{n,00}^2$ \cite{Burkey:2004, Park:2006}. 

\begin{figure}[bt]
\includegraphics[width=.985\textwidth]{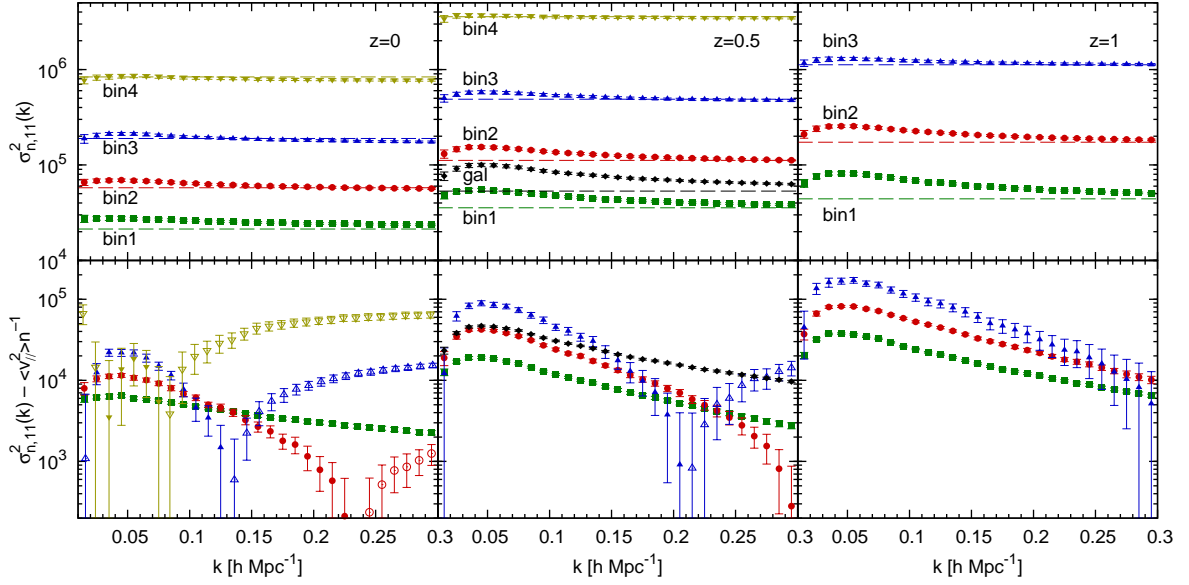}
\caption{{\it Top}: Shot noise power spectrum for momentum power spectrum $\sigma_{n,11}$ measured using the matter-halo cross power spectra.
The horizontal lines with the same color are the shot noise predicted with the Poisson model, $\langle v_\parallel^2\rangle/n$.
{\it Bottom}: The difference between the measurement and the Poisson model. 
The filled and open points show positive and negative values, respectively. 
}
\label{fig:shot}
\end{figure}

In the top panels of figure \ref{fig:shot} we show the shot noise measured for the halo momentum field 
$\sigma_{n,11}^2$. The horizontal lines with the same color as the measured points are the corresponding 
Poisson model. The bottom panels show 
the deviation of the shot noise from the Poisson model, 
$\sigma_{n,11}^2(k)-\langle v_{\parallel}^2 \rangle/n$. 
As is clear from the figure, the behavior of the shot noise measured from simulations differs among 
halo subsamples with different mass, and it changes a lot as a function of scale. 
It is expected to some extent, because even though equation (\ref{eq:shot_00}) is the true expression for the non-Poissonian shot noise for the halo density spectrum, that for the halo momentum spectrum (equation \ref{eq:shot_11}) is just the analogy and holds only in the limited case where the momentum field is uncorrelated with stochasticity. 
Moreover, we have tested if adopting equation (\ref{eq:shot_11}) improved the accuracy of modeling $P_{11}^{r,h}$ and found that the results were not improved very much compared to the case with the Poisson shot noise. For simplicity, we thus assume in the following analysis the Poisson model to subtract the shot noise from the measured auto power spectrum of halo momentum fields $P^{r,h}_{11}$ and $P^{s,h}_{11}$.

The resulting auto power spectra of halos for mass bin1 and bin3 in real space $P^{r,h}_{11}$ at $z=0$ 
after the shot noise is subtracted are shown respectively at the middle and right panels in
the third row of figure \ref{fig:pkmu_01_11}. 
The amplitude of $P^{r,h}_{11}$ exceeds linear theory at large scales similarly to the case of $P^{r,h}_{01}$. 
The behavior of $P^{r,h}_{11}$ is well explained by nonlinear PT for halos with the lowest mass, while there are systematic deviations for the result with more massive halos. 
It is most likely because the assumption of the Poisson model for shot noise $\langle v_\parallel^2\rangle/n$ is incorrect when the number density $n$ is small. 
Because exploring the shot noise on the momentum power spectrum is not the main purpose of this paper, the accuracy of our model for the real-space spectra of momentum field with different halo mass is presented in appendix \ref{sec:real_space}. The effect of shot noise on the momentum power spectrum needs to be studied in detail when analyzing real surveys. 

\begin{figure}[bt]
\includegraphics[width=.985\textwidth]{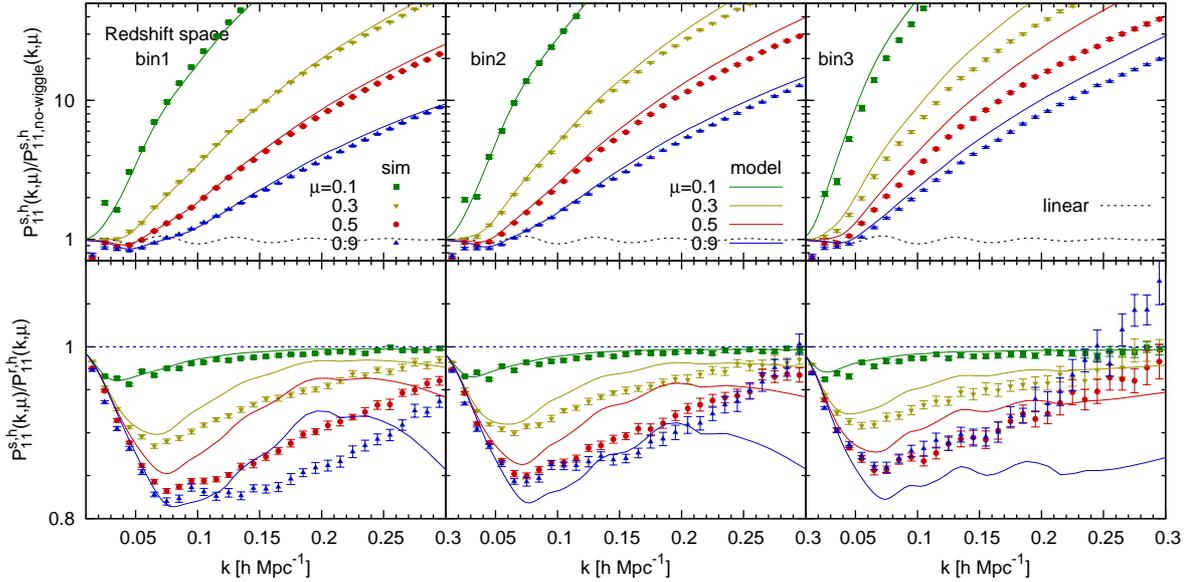}
\caption{
{\it Top row}: Momentum auto power spectra for halos in redshift space
 divided by their no-wiggle spectra at $z=0$, $P_{11}^{s,h}/P_{11,\text{no-wiggle}}^{s,h}$.
{\it Bottom row}: Ratios of the spectra in redshift space and real space, $P_{11}^{s,h}(k,\mu)/P_{11}^{r,h}(k,\mu)$.
The corresponding linear theory predictions are shown as the dotted line with the same color 
while our nonlinear PT predictions are as the solid lines.
}
\label{fig:pkmu_11_z007_3}
\end{figure}

\begin{figure}[bt]
\includegraphics[width=.985\textwidth]{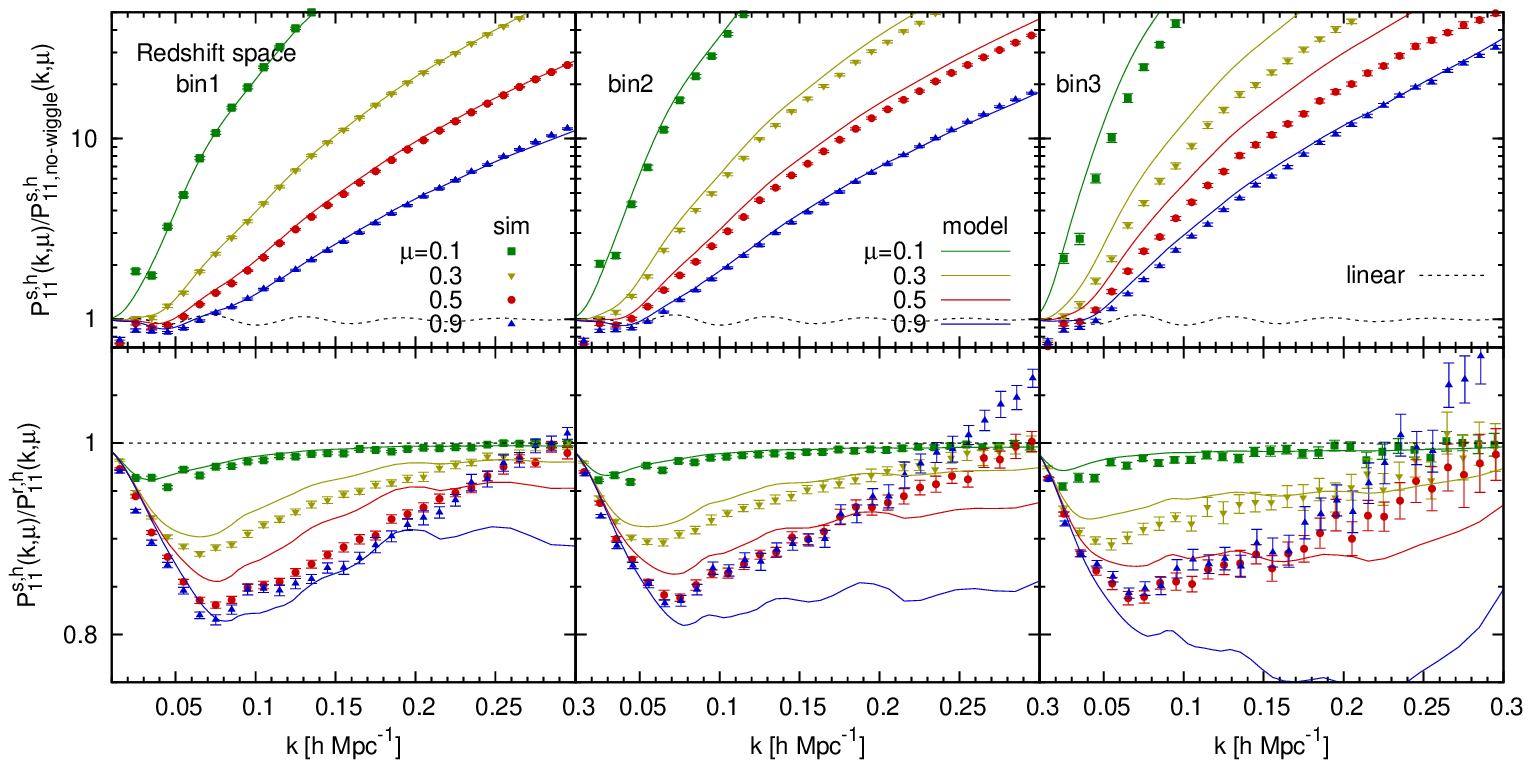}
\caption{Same as figure \ref{fig:pkmu_11_z007_3} but for $z=0.5$.  }
\label{fig:pkmu_11_z005_3}
\end{figure}

\begin{figure}[bt]
\includegraphics[width=.985\textwidth]{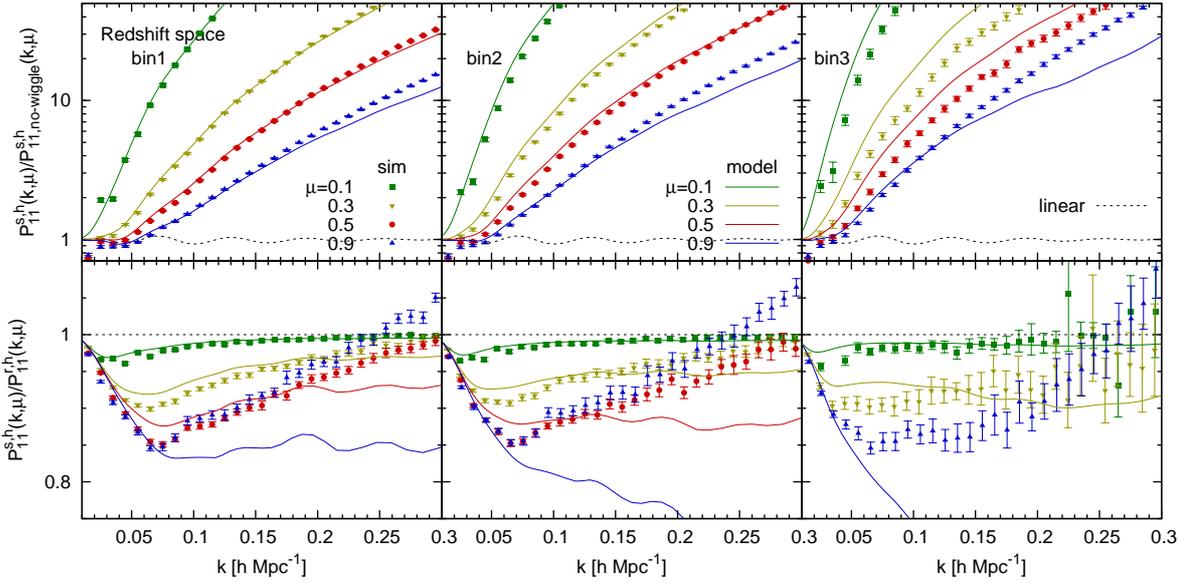}
\caption{Same as figure \ref{fig:pkmu_11_z007_3} but for $z=1$. }
\label{fig:pkmu_11_z004_3}
\end{figure}

Finally we present the auto power spectrum of halos in redshift space  $P^{s,h}_{11}(k,\mu)$.
The resulting redshift-space spectra with the shot noise subtracted for mass bin1 and bin3 
at $z=0$ are shown at the middle and right panels in the bottom row of figure \ref{fig:pkmu_01_11}. 
The nonlinear PT prediction for halos in redshift space achieve almost the same precision compared to the real-space results, and has better agreement to $N$-body results for halos than for dark matter. 
The ratios of these redshift-space momentum spectra to the corresponding linear no-wiggle spectra are presented at the second rows in figures \ref{fig:pkmu_11_z007_3}, \ref{fig:pkmu_11_z005_3} and \ref{fig:pkmu_11_z004_3} 
for $z=0$, 0.5, and 1, respectively. 
Overall, our nonlinear PT at one-loop level for the momentum power spectra of halos in redshift space 
$P^{s,h}_{11}$ is as accurate as that in real space. 
The ratio $P^{s,h}_{11}/P^{r,h}_{11}$ is shown at the bottom rows of figures \ref{fig:pkmu_11_z007_3}, \ref{fig:pkmu_11_z005_3} and \ref{fig:pkmu_11_z004_3}. 
The ratio is closer to unity, linear theory prediction, than that for dark matter. 
The deviation from unity which is caused by fully nonlinear RSD and expected to be small is 
still at $\sim 20\%$ levels. In previous theoretical studies this effect has not been studied, but it needs to be 
taken into account. Our theoretical prediction is well able to predict the $\sim 20\%$ level effect. 
In order to improve the prediction up to smaller scales, however, we need to model the shot noise on 
$P^{s,h}_{11}$ more correctly than the Poisson model.


\section{Configuration-space analysis}\label{sec:configuration_analysis}

Observationally, peculiar velocity statistics have been studied more intensively in configuration space than in Fourier space. In this section we study two basic velocity statistics in configuration space: the mean infall velocity and velocity auto correlation function, that are respectively related to the density-momentum cross power spectrum and momentum auto power spectrum in Fourier space which we have studied so far. We define the Fourier transform of the power spectra of the density-weighted velocity moments as
\be
 \xi_{LL'}^X(\vr) + \delta_{0L}\delta_{0L'}= \left\langle T_\parallel^{X,L}(\vx_b) T_\parallel^{X,L'}(\vx_a) \right\rangle 
 = \int \frac{d^3k}{(2\pi)^3} P_{LL'}^X(k,\mu)e^{-i{\bf k}\cdot{\bf r}}, \label{eq:xiLL}
\ee
where $\vr=\vx_b-\vx_a$, $T_\parallel^L(\vx)$ is defined in equation (\ref{eq:vel_moments}), $\delta_{LL'}$ is the Kronecker delta, and the superscript $X$ is $r$ for real pace and $s$ for redshift space. The functions $\xi_{00}^r$ and $\xi_{00}^s$ are the well-known density correlation functions for a given sample in real space and in redshift space, respectively, and simply written as $\xi_{00}^X=\xi^X$.
In this section we present the velocity statistics in real space and redshift space, and relate them to 
the correlation functions of the velocity moments defined above. 
We also derive the contribution from RSD to the Kaiser limit for the mean infall velocity in redshift space. 


\subsection{Mean pairwise infall momentum}

The mean streaming velocity between pairs of galaxies, 
${\bf v}_{\rm mean}$, was originally introduced by \cite{Davis:1977} in the context of the BBGKY theory. 
Because only the line-of-sight component of velocity, $v_\parallel(\vx)=\vv(\vx)\cdot \hat{\vz}$, is measurable 
in observations, we consider the mean streaming velocity along the line of sight. 
The density-weighted, line-of-sight pairwise velocity for both dark matter and halos is described as
\cite{Davis:1977, Fisher:1995, Scoccimarro:2004};
\be
\left\langle \left[1+\delta(\vx_a)\right]\left[1+\delta(\vx_b)\right]
\left[v_\parallel(\vx_b)-v_\parallel(\vx_a)\right] \right\rangle 
=-2\left\langle \delta(\vx_b)p_\parallel(\vx_a)\right\rangle
=-2\xi_{01}^r(\vr), \label{eq:mean_vel} 
\ee
where $p_\parallel(\vx)$ is the momentum density field given in equation (\ref{eq:momentum}). 
As in section \ref{sec:theory}, a quantity with superscripts $r$ is defined in real space, distinguished from redshift-space quantities denoted with $s$ that we will discuss in the following, and 
the superscripts $m$ and $h$ are omitted when a given equation holds both for dark matter and halos. 
Since the mean infall velocity in configuration space is related to the density-momentum cross power spectrum through equations (\ref{eq:xiLL}) and (\ref{eq:mean_vel}), $\xi_{01}^r$ has only the dipole component, $\xi_{01}^r(\vr)=\xi_{01,1}^r(r) \mu$.
Note that  the mean pairwise velocity in equation (\ref{eq:mean_vel}) is velocity weighted by 
density, thus the mean pairwise momentum. 
It is related to the commonly used, pair-weighted velocity through $-2\xi_{01}^r=[1+\xi^r]v_{\parallel{\rm mean}}^r$, where 
$v_{\parallel{\rm mean}}^r$ is the radial component of the pair-weighted infall velocity, $v_{\parallel{\rm mean}}={\bf v}_{\rm mean}\cdot \hat{z}$. 
We will make a comparison of the density-weighted velocity statistics to pair-weighted velocity statistics in appendix \ref{sec:pair_velocity}.
In linear theory, the equation of the real-space mean streaming is described for halos as
\cite{Davis:1977, Fisher:1995, Juszkiewicz:1999, Sheth:2001, Scoccimarro:2004};
\be
\xi^{r,h}_{01,{\rm lin}}(\vr) =
\mu\frac{f}{2\pi^2b} \int kdk P^{r,h}_{00,{\rm lin}}(k)j_1(kr). \label{eq:mean_vel_lin}
\ee
The linear-theory expression for dark matter is obtained by replacing the superscript $h$ with $m$ and 
substituting $b=1$. 
See \cite{Sheth:2001, Bhattacharya:2008} for nonlinear prediction of the mean streaming velocity 
based on the halo model and \cite{Reid:2011a}  for prediction based on Lagrangian perturbation theory \cite{Matsubara:2008a}.
  
In redshift space, where the radial velocity components can be measured in peculiar velocity surveys, 
we can also write down the mean streaming momentum in the same way, as 
\be
\left\langle \left[1+\delta^s(\vx_a)\right]\left[1+\delta^s(\vx_b)\right]\left[v_\parallel^s(\vx_b)-v_\parallel^s(\vx_a)\right] \right\rangle 
=-2\left\langle \delta^s(\vx_b)p_\parallel^s(\vx_a)\right\rangle = -2\xi^{s}_{01}(\vr). \label{eq:mean_vel_red} 
\ee
Note that although the velocity terms in the rhs of equations (\ref{eq:mean_vel}) and (\ref{eq:mean_vel_red}) 
look the same, $\vr=\vx_b-\vx_a$ in equation (\ref{eq:mean_vel_red}) is the separation 
measured in redshift space. 
Using equations (\ref{eq:p_ll_ang}) and (\ref{eq:xiLL}), 
we can derive full angular dependence of $\xi^s_{01}(\vx)$ as 
\be
\xi^s_{01}(\vr)=\sum_{j=1,3,\cdots} \xi^s_{01,\mu^j}(r)\mu^{j}. \label{eq:xi_01_ang}
\ee

At the linear theory limit for $\xi^s_{01}$, 
we have additional anisotropic terms compared to $\xi^r_{01, {\rm lin}}$ due to linear RSD as follows: 
\bey
\xi^{s,h}_{01,{\rm lin}}(\vr)
&=&\left\langle \delta^{s,h}(\vx_b)v_\parallel^{s,h}(\vx_a) \right\rangle  \nonumber \\
&=& if\int\frac{kdk}{(2\pi)^3} P^{r,m}_{00,{\rm lin}}(k)\int d{\cal O}_\vk e^{-i\vk\cdot\vr}  (b+f\mu_{lk}^2)\mu_{kx} \nonumber \\
&=&\left\langle \delta^h(\vx_b)v_\parallel^h(\vx_a) \right\rangle 
	+if^2\int\frac{kdk}{(2\pi)^3}P^{r,m}_{00,{\rm lin}}\int d{\cal O}_\vk e^{-kr\mu_{kr}} \mu^3_{lk} \nonumber \\
&=& \xi^{r,h}_{01,{\rm lin}}(\vr) 
	+ if^2\int\frac{kdk}{(2\pi)^3}P^{r,m}_{00,{\rm lin}}
	\int d{\cal O}_\vk e^{-kr\mu_{kr}}\left(\frac{2}{5}{\cal P}_3(\mu_{lk}) +\frac{3}{5}{\cal P}_1(\mu_{lk}) \right) \nonumber \\
&=& \left(1+\frac{3f}{5b}\right)\xi^{r,h}_{01,{\rm lin}}(\vr) - \left(\frac{3}{5}\mu-\mu^3\right)\frac{f^2}{2\pi^2}\int kdk P_{00,{\rm lin}}^{r,m}(k)j_3(kr), \label{eq:xi01_lin} 
\eey
where $\mu_{lk}=\hat{\vz}\cdot\hat{\vk}$ and $\mu_{kx}=\hat{\vk}\cdot\hat{\vx}$, and $\mu_{lk}=\mu_{kx}$ since 
we assume the distant-observer approximation. There are thus $\xi^s_{01,\mu^1}$ and $\xi^s_{01,\mu^3}$ in equation (\ref{eq:xi_01_ang}) that have linear order contributions. 

We measure the velocity statistics from $N$-body simulations using the direct pair counting in configuration space. 
Since only the line-of-sight component of velocity is observable in peculiar velocity 
surveys, an estimator for the mean pairwise velocity in real space using only the observables was proposed by \cite{Davis:1996, Ferreira:1999}, as
\be
v_{\rm mean}^X(r)= \frac{\sum_{i,j}{(\vv_i^X \cdot \hat{\vx}_i - \vv_j^X\cdot \hat{\vx}_j) c_{ij} }}
{\sum_{i,j}{c_{ij}^2}}, \label{eq:mean_vel_est}
\ee
where the superscript $X=r$ or $s$ and 
\be 
c_{ij} = {\bf \hat{r}}\cdot \frac{{\bf \hat{x}_{i}} + {\bf \hat{x}_{j}}}{2}.
\ee
$c_{ij}$ is simply equal to $\mu$ under the distant observer approximation. 
This estimator was adopted by several observational and numerical studies 
\cite[e.g.,][]{Juszkiewicz:2000, Bhattacharya:2008} and it is based on mean pair-weighted velocity. 
Throughout this paper except in appendix \ref{sec:pair_velocity}, we analyze the density-weighted velocity in this paper, thus we adopt the following estimator \cite[e.g.,][]{Sheth:2009}, 
\be
-2\xi_{01}^X(r)=\left[1+\xi^X(r)\right] v_{\rm mean}^X(r)=
\frac{\sum_{i,j}{(\vv_i^X \cdot \hat{\vx}_i - \vv_j^X \cdot \hat{\vx}_j) c_{ij} }}
{\sum_{i,j}{c_{ij,{\rm rand}}^2}}, \label{eq:mean_vel_est_mas}
\ee
where $\sum_{i,j}{c_{ij,{\rm rand}}^2}=\left[1+\xi^{X}(r)\right]^{-1}\sum_{i,j}{c_{ij}^2}$, and 
$c_{ij,{\rm rand}}$ is the same as $c_{ij}$ but without clustering. 
This estimator gives the dipole component of the mean pairwise momentum. As we often do for the monopole component for the density correlation function, we omit the order of $l=1$ for the dipole, thus $\xi^{X}_{01}(r)\equiv \xi^X_{01,1}(r)=3\int^1_0 \xi^{X}_{01}(\vr){\cal P}_1(\mu)d\mu$. In real space it is simply $\xi^{r}_{01}(r)=\xi^{r}_{01,\mu^1}$, while in redshift space we have 
$\xi^{s}_{01}(r)=\xi^{s}_{01,\mu^1}+\frac{3}{5}\xi^{s}_{01,\mu^3}+\frac{3}{7} \xi^{s}_{01,\mu^5}+\cdots$.

\begin{figure}[bt]
\includegraphics[width=.985\textwidth]{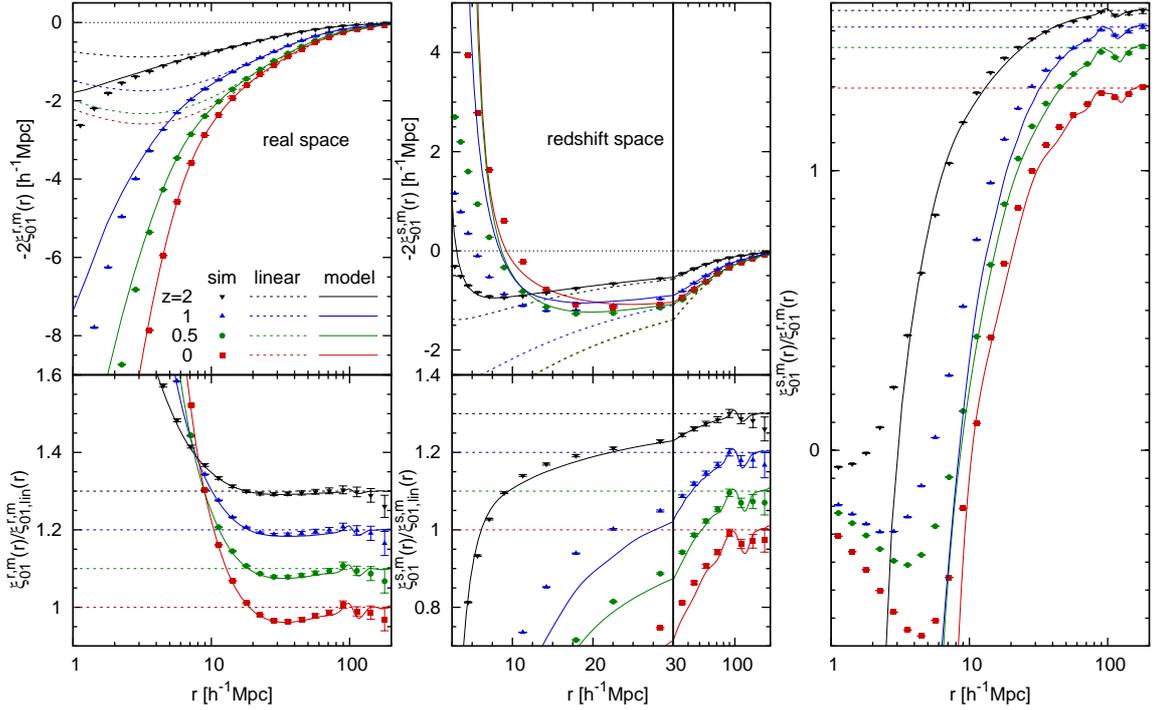}
\caption{{\it Upper panel of left set}: Mean pairwise momentum of dark matter in real space $-2\xi_{01}^{r,m}(r)$. The dotted lines with the same color are the corresponding linear theory predictions, while the solid lines show the Fourier transform of $P_{01}^{r,m}$ measured from simulations. 
{\it Lower panel of left set}: the ratios of the mean pairwise momentum to linear theory. For clarity, the results are offset by $+30\%$, $+20\%$, $+10\%$ and $0\%$ from $z=2$ to $z=0$. {\it Middle set}: same as the left set but results for the dipoles in redshift space $-2\xi_{01}^{s,m}(r)$. The solid lines are our model predictions based on nonlinear PT. To clearly show both the upturn at small scales and large-scale asymptotic behavior,   mixed linear and logarithmic scales are used for the horizontal axis. 
  {\it Right panel}: Ratio of mean pairwise momentum in redshift space to that in real space. The dotted lines show linear theory predictions $1+3f/5$ while the solid lines are nonlinear PT results.}
\label{fig:p_pair_dm}
\end{figure}


\subsubsection{Dark matter}
In the upper left panel of figure \ref{fig:p_pair_dm}, we show the angularly-averaged, dipole component of mean pairwise momentum for dark matter in real space $-2\xi_{01}^{r,m}$ measured using the estimator (equation \ref{eq:mean_vel_est_mas}) at $z=0$, 0.5, 1 and 2. 
Note that the velocity field is normalized by ${\cal H}$ as was stated in section \ref{sec:sim}, so the mean pairwise momentum thus has a unit of \himpc. Because two objects moving towards each other has negative contribution, the sign of $-2\xi^{r,m}_{01}(r)$ is negative at all scales. At all the four redshifts linear theory predictions, shown as the dotted lines, explain the trend of the $N$-body results up to $\sim 15\himpc$. In the lower left panel of figure \ref{fig:p_pair_dm}, these results divided by the corresponding linear theory are plotted. For clarity the results are offset by $+30\%$, $+20\%$, and $+10\%$ for results at $z=2$, 1 and 0.5, respectively. A careful look at the results reveals that the linear theory prediction for the measured mean infall momentum starts to deviate from the $N$-body results at larger scales for lower redshifts, as previously found \cite[e.g.,][]{Bhattacharya:2008, Reid:2011a}. In the both panels of the left set, the solid lines show the Fourier transform of $-2P_{01}^{r,m}$ measured from simulations. 

At the upper panel of the middle set of figure \ref{fig:p_pair_dm} we show the dipole moment of the redshift-space mean pairwise momentum. The change of sign for $-2\xi^{s,m}_{01}(r)$ occurred at $r\sim 10\himpc$ was seen in the recent observation \cite{Hand:2012}. This upturn can be seen in linear theory, but the scale of the turn comes to much smaller scales and the change is also small. On the other hand, the upturn is precisely predicted by nonlinear PT shown as the solid line. Note that while the amplitude of the mean infall momentum monotonically increases as redshift decreases in real space, it is not the case at low redshift in redshift space. It is because the contribution of RSD to the mean streaming velocity is proportional to $f^2$, not $f$. The ratio of the measured mean infall momentum in redshift space to the corresponding linear theory is presented at the bottom panel of the middle set of figure \ref{fig:p_pair_dm}. The measurement is more suppressed from the linear theory in redshift space than that in real space, because the mean infall momentum in redshift space is affected by nonlinear velocity dispersion. It is interesting to note that the deviation of the measurement in redshift space from linear theory is larger even at large scales, around BAO scales of $\simeq 105\himpc$.
These trends in redshift space are very well captured by the nonlinear PT. 
At the right panel in figure \ref{fig:p_pair_dm}, the ratios $\xi^{s,m}_{01}(r)/\xi^{r,m}_{01}(r)$ are shown. 
The linear theory predictions, that are $(1+3f/5)$, are shown as the horizontal lines with the same color. 
As expected the nonlinear PT has good agreement with the $N$-body results. 

\begin{figure}[bt]
\includegraphics[width=.985\textwidth]{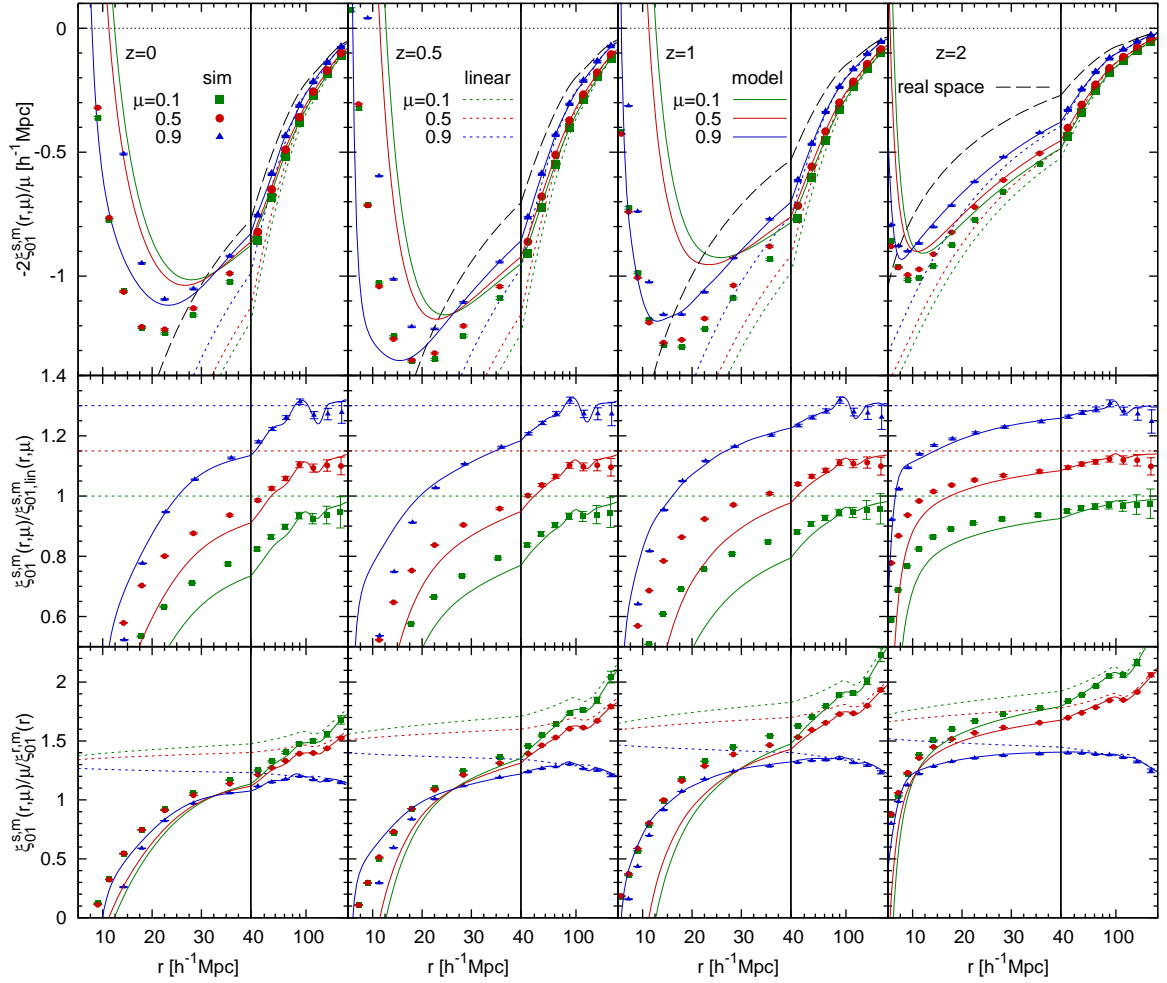}
\caption{
{\it Top row}: mean pairwise infall momenta of dark matter measured in redshift space, $-2\xi^{s,m}_{01}(r,\mu)/\mu$ for the three directions, $\mu=0.1$ (green), $\mu=0.5$ (red) and $\mu=0.9$ (blue). 
The real-space dipole of the mean infall, $-2\xi^{r,m}_{01}(r)$, is also plotted as the black dashed line for comparison. 
The linear theory and nonlinear PT predictions are shown as the dotted and sold lines with the same color, respectively.
{\it Middle row}: the ratio of the redshift-space pairwise velocity to linear theory. For clarity, the results for 
$\mu=0.5$ and $\mu=0.9$ are offset by $+15\%$ and $+30\%$, respectively. 
{\it Bottom row}: the ratio of the mean streaming velocities in redshift space and in real space. 
Note that mixed linear and logarithmic scales are used for the horizontal axis. 
 }
\label{fig:p_pair_2d_dm}
\end{figure}

Unlike the infall velocity in real space, the velocity in redshift space is expressed as the infinite sum 
of odd powers of $\mu$ due to RSD, starting from $\mu^1$, as we 
have seen in equation (\ref{eq:xi_01_ang}) and also in equation (\ref{eq:xi01_lin}) for the 
linear theory limit.
In order to see the anisotropy on the mean pairwise momentum due to RSD, we consider an estimator, 
\be
\frac{-2\xi_{01}^{s}(r,\mu)}{\mu}=
\frac{\sum_{i,j}{(\vv_i^{s} \cdot \hat{\vx}_i - \vv_j^{s} \cdot \hat{\vx}_j) c_{ij} }}
{\sum_{i,j}{c_{ij,{\rm rand}}^2}}. \label{eq:mean_vel_est_mas_2d}
\ee
This is the same estimator as that of equation (\ref{eq:mean_vel_est_mas}),
but the sum at the right-hand side is taken over each $\mu$ bin.

We plot the dark matter pairwise velocity in redshift space, $-2\xi^{s,m}_{01}(r,\mu)/\mu$, 
at the top row in figure \ref{fig:p_pair_2d_dm}.
For comparison we also plot $-2\xi^{r,m}_{01}(r)$, which is the same quantity as in figure \ref{fig:p_pair_dm}.
The amplitude of $\xi^{s,m}_{01}/\mu$ is larger for separation perpendicular to the line-of-sight ($\mu=0$) 
than that along the line-of-sight ($\mu=1$).
The difference between $\xi^{s,m}_{01}(r,\mu=0)$ 
and the real-space one is also larger than that between 
$\xi^{s,m}_{01}(r,\mu=1)$ and the real-space one, opposite of our expectation in Fourier space. 
The $N$-body results of $\xi^{s,m}_{01}$ for the separation along the line-of-sight at the upturn scales are better predicted by our nonlinear PT than those perpendicular to the line-of-sight. 
In the second row of figure \ref{fig:p_pair_2d_dm}, the ratio of the 2D pairwise streaming velocity in 
redshift space to the corresponding linear theory prediction, 
$\xi^{s,m}_{01}(r,\mu)/\xi^{s,m}_{01,{\rm lin}}(r,\mu)$ is shown.
The deviation from linear theory is larger for $\xi^{s,m}_{01}$ at $\mu=0$ than for $\mu=1$.
At $z=0$ the result for $\mu=0.9$ is consistent with linear theory at $r\geq 80\himpc$ 
while that for $\mu=0.1$ agrees with linear theory at $r > 100\himpc$. 
While our PT prediction is in good agreement with the $N$-body measurements for larger $\mu$, 
the prediction for the transverse separation is less accurately agreed with the measurement. 
In the bottom row of figure \ref{fig:p_pair_2d_dm}, we show the anisotropic pairwise streaming 
velocity in redshift space divided by the isotropic dipole in real space, $\xi^{s,m}_{01}(r,\mu)/\mu/\xi^{r,m}_{01}(r)$. Because the anisotropy is 
characterized by $f(z)$, the higher-redshift results are more anisotropic. 
Again one can see the significant deviation from linear theory for the mean pairwise momentum at 
$\mu=0$ even at very large scales. 


\subsubsection{Halos}
Now let us extend the analysis to biased objects, dark matter halos.
The real-space mean pairwise momentum is just a Fourier transform of $P_{01}^r$ that was studied and the $N$-body results were compared to perturbation theory in \cite{Vlah:2013}. However, it is useful to show the configuration-space results because several previous works adopted the configuration-space statistics and also they can also be compared to the corresponding redshift-space statistics we will present below. We thus show the real-space results in appendix \ref{sec:real_space}. 

\begin{figure}[bt]
\includegraphics[width=.985\textwidth]{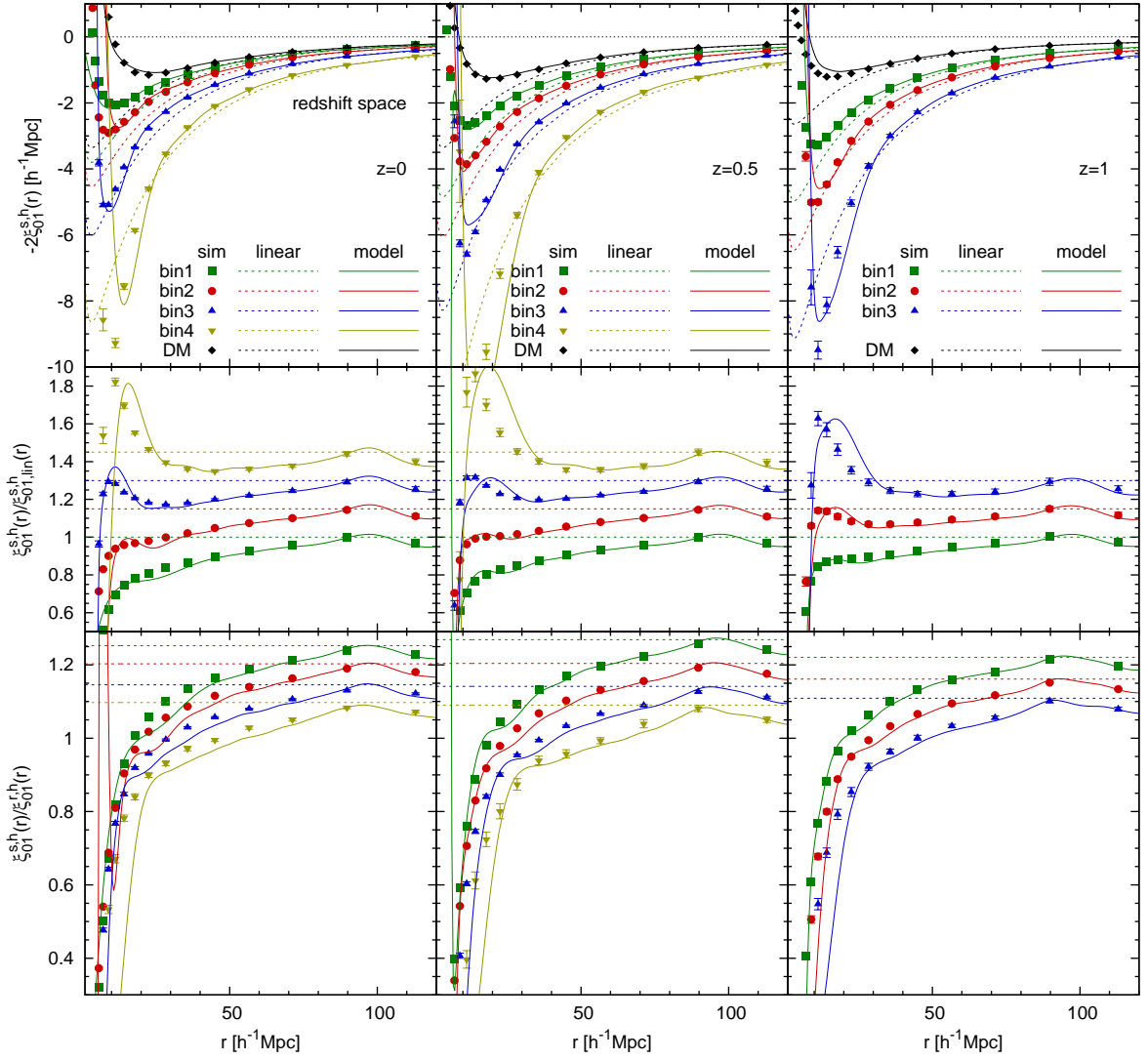}
\caption{{\it Top row}: Mean pairwise infall momenta of halos in redshift space $-2\xi_{01}^{s,h}(r)$. 
  For comparison the results for dark matter are shown as the black points.
  The dotted lines with the same color are the
  corresponding linear theory predictions while the solid lines are the corresponding nonlinear PT predictions. 
  {\it Second row}: The ratios of the
  mean pairwise momenta to the corresponding linear theory. 
  The results are offset by $+0\%$, $+15\%$, $+30\%$ and
  $+45\%$ for halos from the lightest to heaviest mass bins. 
{\it bottom row}: The ratios of the mean pairwise momenta in redshift space and real space, 
$v^{s,h}_{\rm mean}(r)/v^{h}_{\rm mean}(r)$. The linear theory denoted by the dotted horizontal line gives the Kaiser factor $(1+3f/5b)$. 
 }
\label{fig:p_pair_s}
\end{figure}

\begin{figure}[bt]
\includegraphics[width=.985\linewidth]{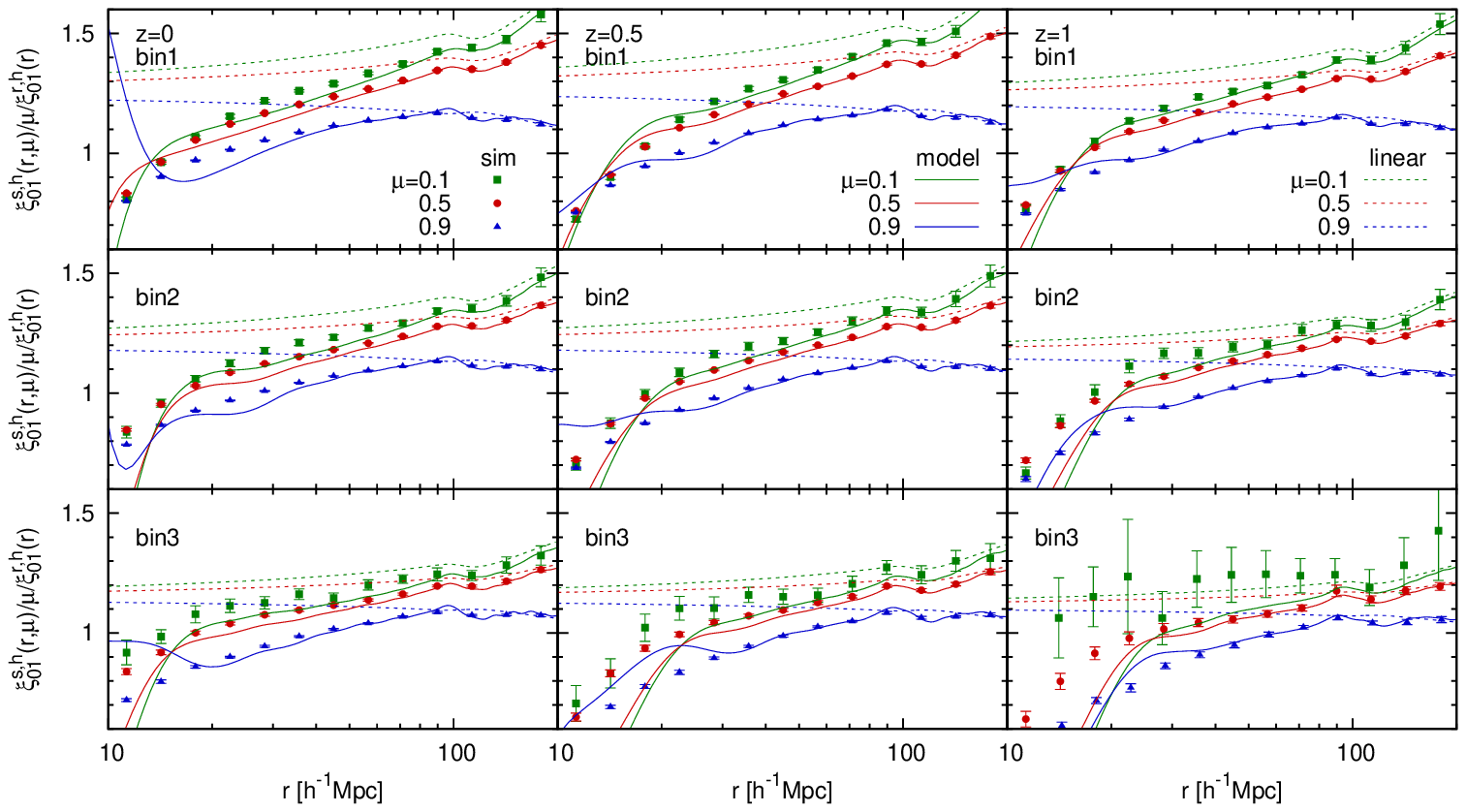}
\caption{The ratios of the mean pairwise momenta of halos in redshift space and the momentum dipole in real space, $\xi^{s,h}_{01}(r,\mu)/\mu/\xi^{r,h}_{01}(r)$. Results for $z=0$, 0.5 and 1 are shown at the left, middle and right columns, respectively. Linear and nonlinear predictions are respectively shown as the dotted and solid lines with the same color as the measurements. 
 }
\label{fig:p_pair_2d_lin}
\end{figure}

We present the pairwise mean infall momenta for halos in redshift space.
At the top row of figure \ref{fig:p_pair_s} we show the angularly-averaged, dipole components.
The results for dark matter, presented already, are also shown for comparison. 
As is the case with dark matter, nonlinear PT at one-loop level can predict well the upturn in 
$-2\xi^{s,h}_{01}$ seen at around $r\sim 10\himpc$. At the second row we show the ratios of $\xi^{s,h}_{01}$ from $N$-body simulations and nonlinear PT to the corresponding liner theory. 
As we have already seen, redshift-space velocity statistics in Fourier space, $P_{01}^s$ and $P_{11}^s$ 
have much more nonlinear contributions than real-space statistics $P_{01}$ and $P_{11}$. 
In configuration space there are significant nonlinear corrections up to very large scales through nonlinear transform from Fourier space \cite{Scoccimarro:2004}. 
Thus the deviation of the redshift-space pairwise momentum $\xi^{s,h}_{01}$ is much more significant at all the scales probed than the real-space one $\xi^{r,h}_{01}$.
At scales $r\leq 100\himpc$ the pairwise momentum for halos starts to be suppressed relative to linear theory due to nonlinear velocity dispersion as is the case for dark matter. 
This large-scale deviation can be well explained by our nonlinear PT prediction. For massive halo subsamples the amplitude of the measured momentum exceeds linear theory at scales of 
$r\sim 30\himpc$ and at smaller scales the amplitude goes below linear theory again. 
This complicated behavior of $\xi^{s,h}_{01}$ is also precisely predicted by nonlinear PT. 
We plot the ratio of the mean pairwise momenta between redshift space and real space at the bottom row in figure \ref{fig:p_pair_s}. 
As in the right panel of figure \ref{fig:p_pair_dm}, the horizontal lines are the linear theory predictions, 
$(1+f(z)/3b_1)$, where $b_1$ is the large-scale constant halo bias (see section \ref{sec:sim}).

In order to see the angular dependence of the mean pairwise infall momentum due to RSD, we 
show in figures \ref{fig:p_pair_2d_lin} the ratio of redshift-space momentum to real space one 
as functions of $(r,\mu)$ measured using the estimator of equation (\ref{eq:mean_vel_est_mas_2d}).
They are compared to the prediction from linear and nonlinear PT, and good agreement of the 
$N$-body results to the latter is found. As we have seen for dark matter in figure \ref{fig:p_pair_2d_dm}, 
linear theory has a poor accuracy also for halos pairwise velocity in redshift space.


\subsection{Momentum correlation function}

The second fundamental velocity statistics in configuration space is an auto-correlation function of density-weighted peculiar velocity \cite{Davis:1977}, the Fourier transform of the momentum auto power spectrum $P_{11}^X$. The momentum correlation is fully described in real space by the $3\times 3$ correlation tensor $\Psi_{ij}^r$, and for a statistically homogeneous and isotropic velocity field, it follows that \cite{Davis:1977, Davis:1983, Gorski:1988} 
\be
\Psi_{ij}^r(\vr) = \left\langle \left[1+\delta(\vx_a)\right]\left[1+\delta(\vx_b)\right]v_{i}(\vx_b) v_{j}(\vx_a)\right\rangle  
= \Psi_\perp(r)\delta_{ij} + \left[ \Psi_\partial(r)-\Psi_\perp(r)\right] \hat{\vr}_i\hat{\vr}_j, \label{eq:vel_corr_def}
\ee
where $\Psi_\partial$ and $\Psi_\perp$ are respectively the correlation functions parallel and perpendicular to 
the line of separation\footnote{Note that the subscript $\parallel$ is usually used in place of $\partial$, as $\Psi_\parallel$, but in this paper $\parallel$ is used to describe a quantity along the line of sight. }.
The statistics for the momentum correlation can be defined using the observable quantities, 
as \cite{Gorski:1989, Groth:1989}
\be
\sum_{i,j}^3\hat{\vz}_{i} \Psi_{ij}^r(\vr) \hat{\vz}_{j} =\left\langle p_\parallel(\vx_b)p_\parallel(\vx_a) \right\rangle 
=\xi_{11}^r(\vr). \label{eq:vel_corr_real_obs}
\ee
Thus, the angular dependence of the momentum correlation function in real space is described as
\be
\xi^r_{11}(\vr)=\xi_{11,\mu^0}^r(r)+\xi_{11,\mu^2}^r(r)\mu^2=\Psi_\perp(r) +\left[\Psi_\partial(r)-\Psi_\perp(r)\right]\mu^2. \label{eq:xi_11_real_ang}
\ee
In linear theory, we have well-known results in real space\cite{Gorski:1988, Peel:2006, Sheth:2009};
\be
\Psi_\perp(r)=\frac{f^2}{2\pi^2}\int dk P_{00}^{r,m}(k)\frac{j_1(kr)}{kr},  \ \ \ \ \ \ 
\Psi_\partial(r)= \frac{f^2}{2\pi^2}\int dk P_{00}^{r,m}(k)j_0(kr) . \label{eq:psi_para}
\ee 

In redshift space we can use exactly the same definitions as in real space (equations \ref{eq:vel_corr_def} and \ref{eq:vel_corr_real_obs}),
so we can write,
\be
\left\langle \left[1+\delta^s(\vx_a)\right]\left[1+\delta^s(\vx_b)\right]v_\parallel^s(\vx_b) v_\parallel^s(\vx_a)\right\rangle   
 =\left\langle p_\parallel^s(\vx_b)p_\parallel^s(\vx_a) \right\rangle 
=\xi_{11}^s(\vr).
\ee
However, because of equation (\ref{eq:xiLL}) in redshift space, 
the redshift-space velocity correlation function is expressed as an infinite sum of the velocity-moment correlation functions,
\be
\xi^s_{11}(\vr)=\sum_{j=0,2,\cdots} \xi_{11,\mu^j}^s(r)\mu^{j}. \label{eq:xi_11_ang}
\ee
As we have seen in the previous section, the velocity correlation in redshift space is equivalent to 
that in real space in the Kaiser limit, $\xi^s_{11,{\rm lin}}(\vr)= \xi^r_{11,{\rm lin}}(\vr)$.

From $N$-body simulations, we measure the velocity correlation function, we use an estimator introduced 
in real space by \cite{Gorski:1989} and modified as, 
\be
\xi^{X}_{11}(r)=\frac{\sum_{i,j}(\vv_i\cdot \hat{\vx}_i)d_{ij}(\vv_j\cdot \hat{\vx}_j)}{\sum_{i,j}d_{ij,{\rm rand}}}, 
\ \ \ \ {\rm where} \ \ \ \ 
d_{ij}=\hat{\vx}_i\cdot\hat{\vx}_j, \label{eq:vel_corr_est_mas}
\ee
$X=r$ or $s$ and $\sum_{i,j}{d_{ij,{\rm rand}}^2}=\left[1+\xi^{X}(r)\right]^{-1}\sum_{i,j}{d_{ij}^2}$.
Under the distant observer approximation, simply $d_{ij}=1$. 
Equation (\ref{eq:vel_corr_est_mas}) gives the angularly-averaged, monopole moment of the velocity correlation function. 
The estimator originally proposed in \cite{Gorski:1989} is pair-weighted, related to our estimator as 
$\psi^{r}(r)=\left[ 1+\xi^r (r)\right]^{-1}\xi^{r}_{11}(r)$, and will be discussed in appendix \ref{sec:pair_velocity}.

\begin{figure}
\includegraphics[width=.985\textwidth]{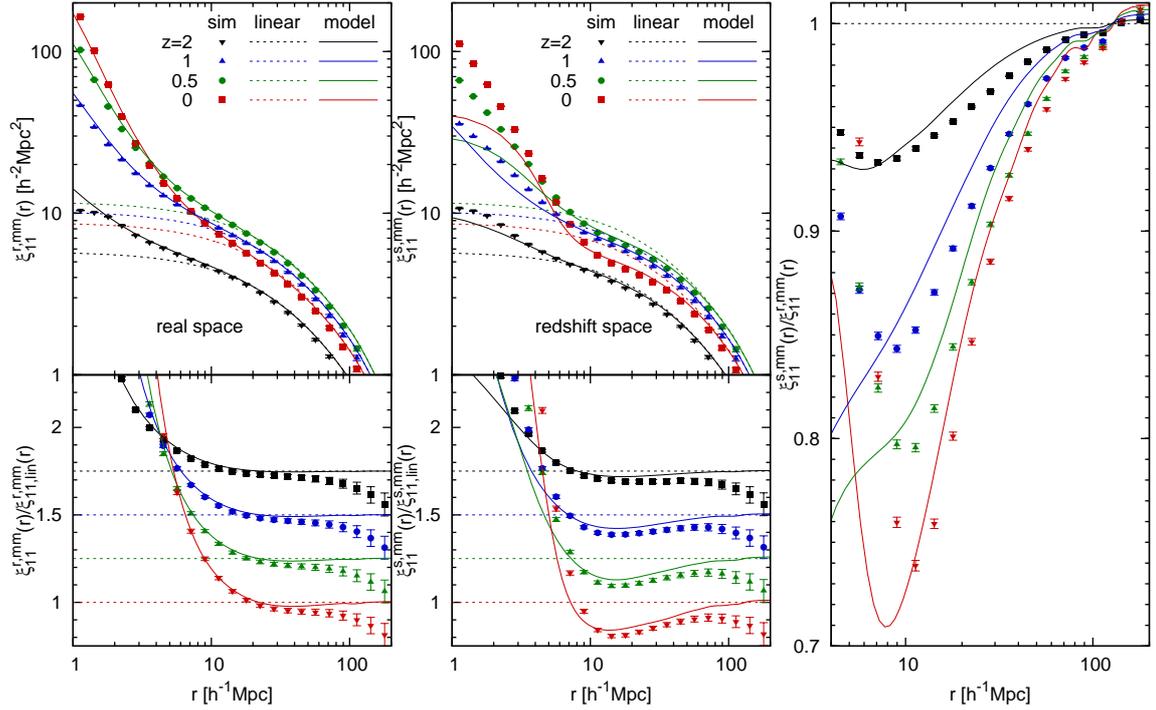}
\caption{{\it Upper panel of left set}: angularly-averaged momentum correlation functions of dark matter in real space
 $\xi^{r,m}_{11}(r)=\langle p_\parallel^m(\vx) p_\parallel^m(\vx+\vr)\rangle$. The dotted lines with the same color are the
  corresponding linear theory predictions, and the solid lines are our nonlinear PT predictions. 
  {\it Lower panel of left set}: the ratios of the
  momentum correlation function to the linear theory. 
  For clarity, the results are offset by $+75\%$, $+50\%$, $+25\%$ and
  $0\%$ for dark matter from $z=2$ to $z=0$. 
 {\it Middle set}: same as left set but 
  for redshift space, $\xi^{s,m}_{11}(r)$. {\it Right panel}: monopole-to-real-space ratio of 
  the momentum correlation functions, $\xi^{s,m}_{11}(r)/\xi^{r,m}_{11}(r)$. 
  The dotted line is the linear theory predictions, which is just unity while the 
  solid lines with the same color as the points are corresponding nonlinear PT predictions.
  }
\label{fig:p_auto_dm}
\end{figure}

\begin{figure}
\includegraphics[width=.985\textwidth]{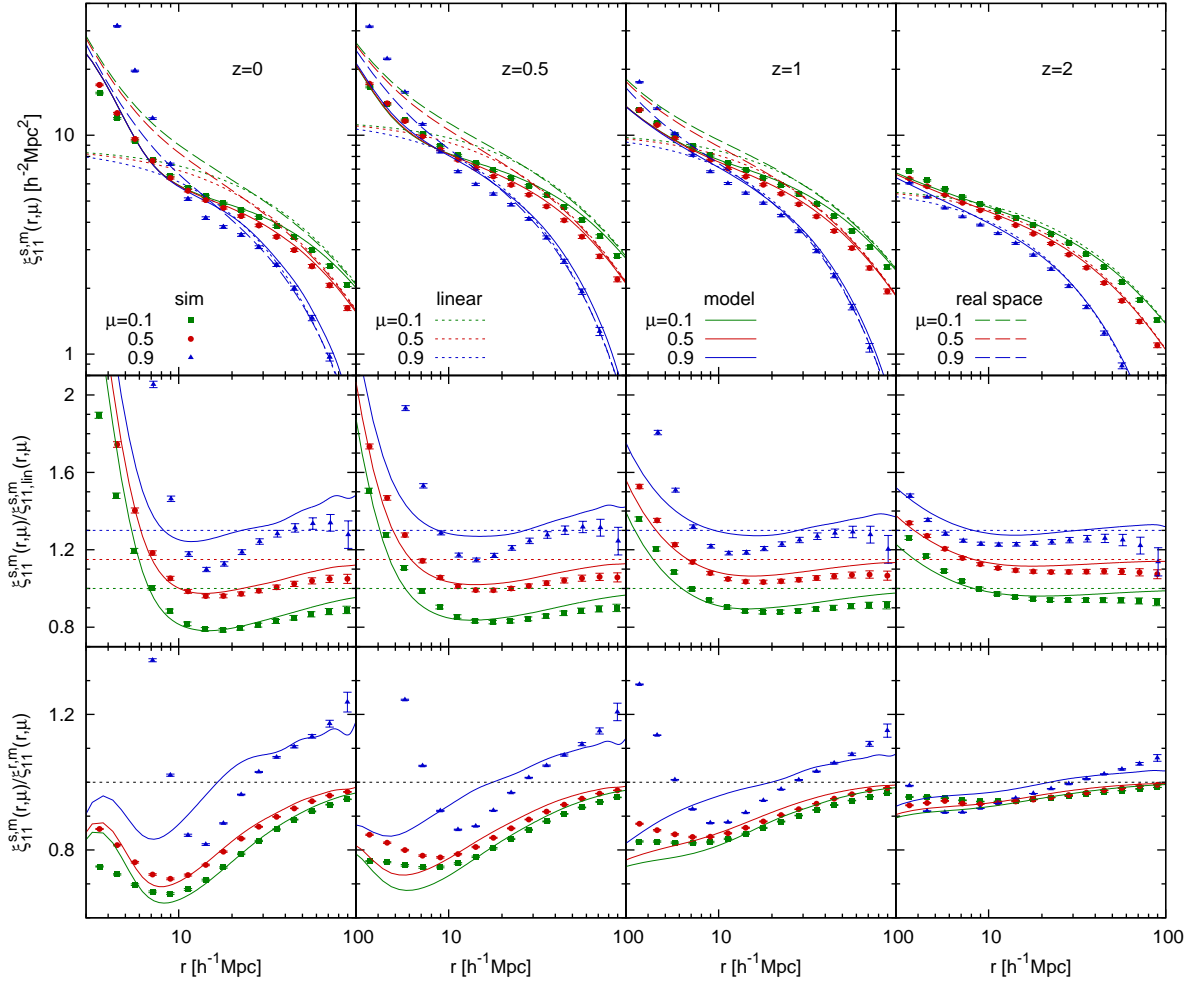}
\caption{
Momentum correlation functions of dark matter in redshift space, $\xi^{s,m}_{11}(r,\mu)$ (first row) 
 for the three angular directions, $\mu=0.1$, 0.5 and 0.9 denoted by the green, red and blue points, respectively. 
 The real-space result $\xi^{r,m}_{11}(r,\mu)$ is shown as the dashed lines for comparison. 
  The second row shows the ratios of the redshift-space momentum correlation function to the linear theory, $\xi^{s,m}_{11}(r,\mu)/\xi^{s,m}_{11,{\rm lin}}(r,\mu)$. 
  The results for $\mu=0.5$ and 0.9 are offset by $+15\%$ and $+30\%$, respectively for clarity. 
  The bottom row shows the ratio of the momentum correlation function in redshift space to that in 
  real space $\xi^{s,m}_{11}(r,\mu)/\xi^{r,m}_{11}(r,\mu)$. 
  In all the panels the dotted and solid lines with the same color as the $N$-body measurements 
  are the corresponding linear theory and nonlinear PT predictions, respectively. 
  }
\label{fig:p_auto_2d_dm}
\end{figure}

\subsubsection{Dark matter}

In the upper left panel of figure \ref{fig:p_auto_dm}, we show the monopole momentum correlation function of dark matter in real space $\xi^{r,m}_{11}(r)$ for $z=0$, 0.5, 1 and 2. The velocity correlation function has a unit of $[\himpc]^2$. Just like the redshift-space mean pairwise momentum, the momentum correlation of dark matter does not simply evolve with redshift because the amplitude depends on $f^2D^2$. $N$-body results start to deviate from linear theory, denoted as the dotted lines, at the scales $r\sim 10\himpc$ for all the four redshifts. At smaller scales measurements from simulations exceed linear theory due to the nonlinear effects. The solid lines are the Fourier transform of our prediction for $P_{11}^{r,h}$: the scalar part of the dark matter $P_{11}^{r,m}$ modeled using $N$-body simulations and the vector part using nonlinear PT. Because the vector part was well modeled using nonlinear PT, the prediction matches the $N$-body result in configuration space up to small scales, $r\sim 5\himpc$ at $z=0$ and $r\sim 3\himpc$ at $z=2$. In the lower left panel of figure \ref{fig:p_auto_dm}, the momentum correlation function divided by the corresponding linear theory is shown. At large scales the simulation results are systematically smaller than linear theory, almost by a constant \cite[e.g.,][]{Bhattacharya:2008, Sheth:2009}. Thus the larger scales we focus on and the smaller the amplitude of the correlation becomes, the lager the deviation of the ratio to linear theory deviates from unity becomes. On such large scales there is a small but non-negligible improvement ($\sim$ a few percent) by applying nonlinear PT. Recently a new method to model the velocity correlation function based on a peculiar velocity decomposition was proposed and tested by \cite{Zhang:2013, Zheng:2013}.

At the upper panel of the middle set in \ref{fig:p_auto_dm}, 
we show the momentum correlation function for dark matter in redshift space, averaged over angle.
Linear theory prediction in redshift space shown as the dotted lines is exactly the same as 
that in real space, while it is not the case for nonlinear PT shown as the solid lines. 
Although the cross power spectra between density and velocity moments $\xi_{0L}^r$ do not contribute to 
$\xi^{s}_{11}$, there are still many terms contributing to it such as $\xi_{12}^r$, $\xi_{13}^r$, $\xi_{22}^r$, and so on. 
Thus the measurements in redshift space start to exceed linear theory at larger scales, particularly for
lower redshifts. Nonlinear PT can explain the measurements up to smaller scales than linear PT does, but 
it also breaks down at larger scales than the case in real space. 
The ratio of the measured momentum correlation function in redshift space to the corresponding linear theory is presented at the bottom panel of the middle set of figure \ref{fig:p_auto_dm}. 
The nonlinear, higher-order terms suppress the amplitude to deviate from linear theory even at $r<90\himpc$ for $z=0$ and $r<50\himpc$ for $z=2$.  One loop PT dramatically improves the agreement with the $N$-body results, up to $r<20\himpc$ for $z=0$ and $r<4\himpc$ for $z=2$. 
At the right panel in figure \ref{fig:p_auto_dm}, the ratios $\xi^{s,m}_{11}(r)/\xi^{r,m}_{11}(r)$ are shown. 
As is the case in Fourier-space analysis, in linear theory the ratio is just unity, thus the deviations from 
unity are entirely from a nonlinear RSD effect. The effect alters the amplitude by at most $\sim 25\%$ for 
$z=0$ and $\sim 7\%$ for $z=2$. Due to the systematic deviations of $\xi^{s,m}_{11}$ and $\xi^{r,m}_{11}$ 
from the linear theory at large scales, there are slight discrepancies in the ratios 
between $N$-body results and nonlinear PT at all scales. However, overall behaviors of the nonlinear RSD are well explained using our one-loop PT.

We show the real-space momentum correlation function of dark matter in 2D,
$\xi^{r,m}_{11}(r,\mu)$, at the first row of figure \ref{fig:p_auto_2d_dm} as the dashed lines. 
We do not show the real-space result at $z=2$ because it is very close to that in redshift space, as shown at the bottom right panel. The results are well explained by linear theory, 
up to $\sim 20\himpc$ at $z=0$ and $\sim 10\himpc$ at $z=2$, although there are small but systematic 
deviations for $\mu=0.9$ at larger scales at lower redshifts. 
The first row of figure \ref{fig:p_auto_2d_dm} also shows the redshift-space momentum correlation function 
$\xi^{s,m}_{11}(r,\mu)$. Its linear theory prediction is equivalent to that in real space. 
Since in practice there are contributions from higher powers of $\mu$ due to RSD, 
the validity of linear theory is limited to very large scales,  
while our nonlinear PT at one loop level captures the deviation of the measurements from linear theory at $r\leq 10\himpc$. 
However, as shown in the second row of figure \ref{fig:p_auto_2d_dm}, 
there is still small disagreement between $N$-body results for $\mu=0.9$ and their nonlinear PT prediction.
The bottom row of figure \ref{fig:p_auto_2d_dm} plots  
the ratio between momentum correlation functions in redshift space and in real space. 
Just like the case of Fourier space in figure \ref{fig:pkmu_11_dm}, the deviation of the ratio from unity 
is caused by nonlinear RSD. 
At $r>10\himpc$ the amplitude is suppressed by $\sim 30\%$ at $z=0$ and $\sim 10\%$ at $z=2$.

\begin{figure}[bt]
\subfigure{\includegraphics[width=.985\textwidth]{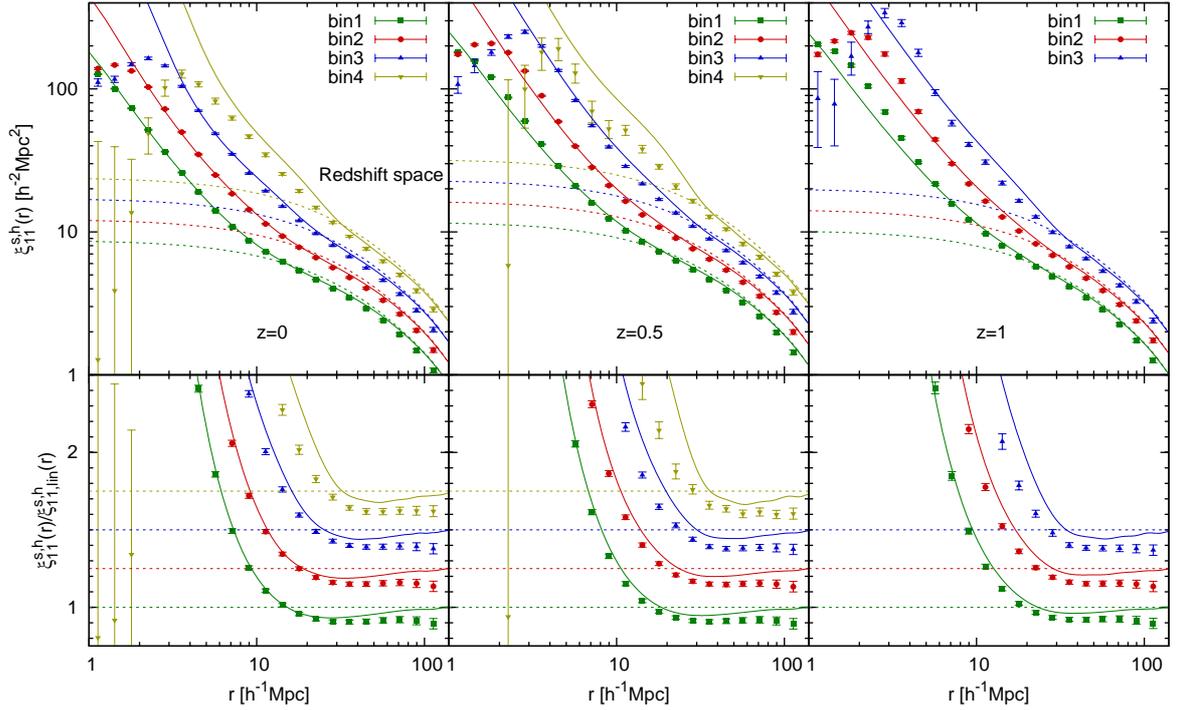}}
\caption{Same as \ref{fig:p_auto_r} but here we show the halo results in redshift space, 
$\xi^{s,h}_{11}(r)$.}
\label{fig:p_auto_s}
\end{figure}

\begin{figure}[bt]
\subfigure{\includegraphics[width=.985\textwidth]{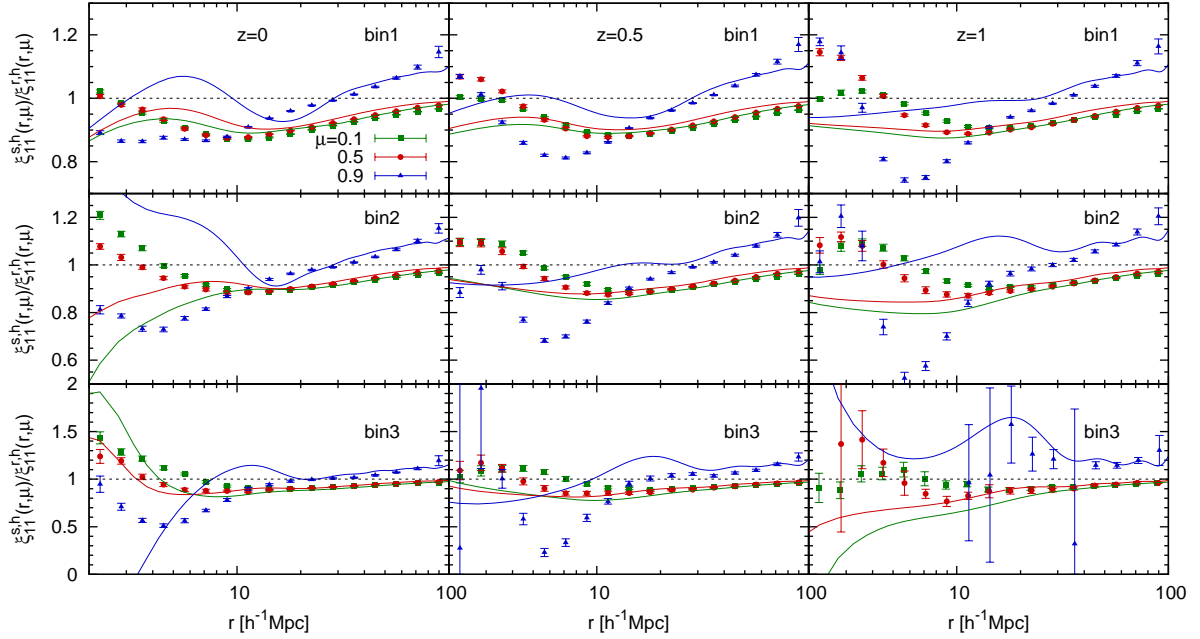}}
\caption{The ratios 
of the momentum correlation functions of halos in redshift space and in real space in 2D, 
$\xi^{s,h}_{11}(r,\mu)/\xi^{r,h}_{11}(r,\mu)$. 
Linear and nonlinear predictions are respectively shown as the dotted (unity) and solid lines with the same color as the measurements.  }
\label{fig:p_auto_2d}
\end{figure}


\subsubsection{Halos}
Finally, let us extend the dark matter analysis to halos.
We discuss the momentum correlation function of halos $\xi_{11}^{r,h}$ with different mass bins at appendix \ref{sec:real_space}. 
At the top row of figure \ref{fig:p_auto_s} we show the angularly-averaged momentum correlation functions 
for halos in redshift space.
Unlike the case for the mean infall momentum, the shape of the momentum correlation function in redshift space is 
very similar to that in real space for all scales proved. 
It implies that even the redshift-space momentum correlation at nonlinear scales is dominated by the 
real-space one that only contains linear-order contributions in the expansion (equation \ref{eq:P11ss}). 
Also the accuracy of nonlinear PT at small scales in redshift space is almost the same as for that in real space. 
The bottom row of figure \ref{fig:p_auto_s} compares $N$-body results to 
linear theory and our nonlinear PT.
Contrary to the real-space case, $N$-body measurements do not agree with linear theory even up to 
very large scales, $\sim 70\himpc$ for the halos with the smallest mass and $\sim 100\himpc$ at 
$z=0$. On the other hand, nonlinear PT agrees with the $N$-body results quite well. 
It is thus necessary to take into account the nonlinear corrections to theoretical models
in order to predict the momentum correlation function in redshift space.

Figure \ref{fig:p_auto_2d} shows the ratio of the redshift-space and real-space momentum correlation functions 
in 2D, $\xi^{s,h}_{11}(r,\mu)/\mu/\xi^{r,h}_{11}(r)$. 
The ratio measured from simulations is compared to linear theory that is unity and to nonlinear PT. 
The deviation of the ratio from unity due to nonlinear RSD is smaller for halos than
for dark matter we have seen in figure \ref{fig:p_auto_2d_dm}, 
and becomes smaller for more massive halos. 
Nonlinear PT agrees with the $N$-body results up to $r\sim 10\himpc$ for $\mu=0.1$ and 0.5 and 
up to $r\sim 30\himpc$ for $\mu=0.9$.


\section{Effects of satellite galaxies on velocity statistics}\label{sec:satellite}
So far we have focused on modeling velocity statistics for dark matter and halos using nonlinear PT. 
In real surveys we observe galaxies that have two types of contributions; central galaxies which 
can be considered to move together with the host halos \cite[but see,][]{Hikage:2013a, Hikage:2013}, 
and satellite galaxies that mainly exist in massive halos. 
Modeling the latter part is not a trivial task, which makes it hard to model the statistics of a galaxy sample. 
It is thus meaningful to test whether the velocity statistics we considered above can be modeled for 
galaxies using PT of halos. 

As a mock galaxy catalog, we use a halo occupation distribution (HOD) modeling which populates dark matter halos with
galaxies according to the halo mass \cite[e.g.,][]{Seljak:2000, Scoccimarro:2001, Cooray:2002, Zheng:2005}. We consider as a mock galaxy sample the galaxies from the Baryon Oscillation Spectroscopic Survey (BOSS) \cite{Schlegel:2009, Eisenstein:2011}, referred to as the ``CMASS" galaxy sample. Galaxies are assigned to the halos at $z=0.5$ using the best fit HOD parameters for CMASS galaxies determined by \cite{White:2011}. For halos which contain satellite galaxies, we randomly pick up the same number of dark matter particles to represent the positions and velocities of the satellites. This method was applied in our previous work \cite{Okumura:2012b} and the good agreement with the observation has been confirmed. In the following we will compare the real-space and redshift-space velocity statistics measured from the mock galaxy sample to the nonlinear PT prediction for the halo subsample that has the same bias as the galaxy sample (bin2 at $z=0.5$).

\subsection{Fourier space}
\begin{figure}[bt]
\subfigure{\includegraphics[width=.985\textwidth]{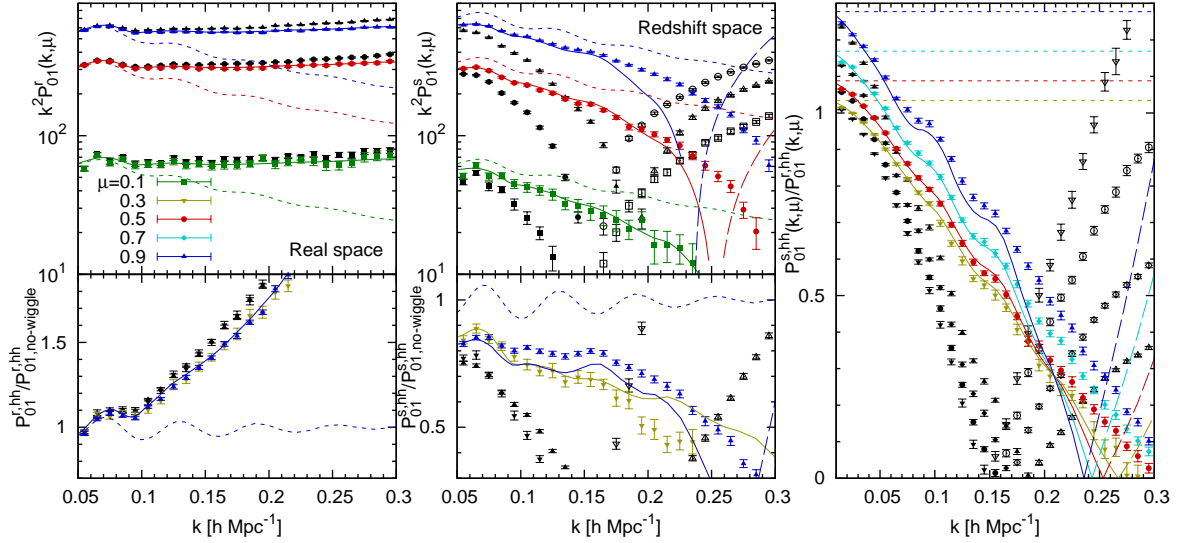}}
\caption{At the upper panel of the left set we show the density-momentum power spectrum in real space $k^2P_{01}^{r,h}$ at $z=0.5$, and at the lower panel we plot its ratio to the corresponding linear theory prediction without BAO, $P_{01}^{r,h}/P_{01,\text{no-wiggle}}^{r,h}$. The middle set is the same as the left set but for the results in redshift space. The right panel shows the ratio of the density-momentum spectra in redshift space and in real space, $P_{01}^{s,h}/P_{01}^{r,h}$. The colored points show the results for halos at $z=0.5$ and the black points with the same point types as the colored points show the results for mock galaxies with the same values of $\mu$. The halo subsample has the same bias as the galaxy sample. The dotted lines show the linear Kaiser model while the solid (positive) and dashed (negative) lines show our one-loop PT prediction. For clarity we show only results for some $\mu$ bins among the five $\mu$ bins. 
 }
\label{fig:pkmu_01_lrg}
\end{figure}

We start the comparison in Fourier space. At the upper panel of the left set in figure \ref{fig:pkmu_01_lrg} we show the density-momentum power spectra in real space $P^{h}_{01}$ of galaxies and halos which have the same bias as the galaxies. The lower panel shows this cross power spectrum divided by the linear theory, $P^{r,h}_{01}/P^{r,h}_{01,{\rm lin}}$. The behavior of $P^{r,h}_{01}$ measured for galaxies in real space is consistent to that for halos to $k\leq 0.14\hmpci$ for higher $\mu$ values. The amplitude of $P^{r,h}_{01}$ for galaxies becomes higher than that for halos at smaller scales, by $\sim 10\%$ at $k\leq 0.2\hmpci$. This behavior has been already seen in \cite{Okumura:2012b}. 

The upper panel of the middle set in figure \ref{fig:pkmu_01_lrg} shows the results for halos and galaxies in redshift space and the lower panel shows their ratios to the corresponding linear theory. The right panel in figure \ref{fig:pkmu_01_lrg} shows the ratio of the redshift-space and real-space cross power spectra. Because of the contributions from satellite galaxies, the galaxy sample has much larger velocity dispersion than halos. Thus the redshift-space density-momentum power spectrum of galaxies is suppressed and changes the sign as is the case for the quadrupole spectrum of galaxy density but at larger scales $k \sim 0.15 \hmpci$ (see figure 5 of \cite{Okumura:2012b}). Obviously, nonlinear PT for halos cannot predict the measured cross spectrum in redshift space except for very large scales because we used linear theory to predict the velocity dispersion of halos. At $k\sim 0.1\hmpci$ the amplitude of $P^{s,h}_{01}$ for galaxies is suppressed relative to nonlinear PT for halos by $\sim 50\%$.

\begin{figure}[bt]
\subfigure{\includegraphics[width=.985\textwidth]{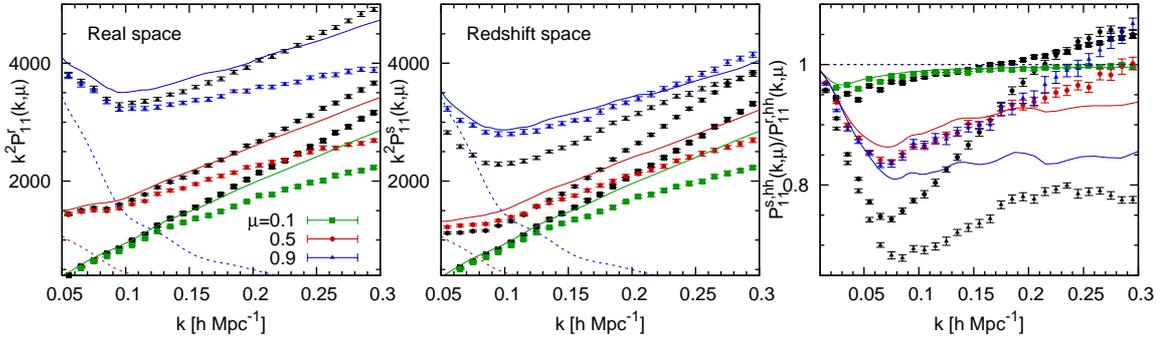}}
\caption{
Same as figure \ref{fig:pkmu_01_lrg} but results for the momentum auto power spectra are shown. The power spectra in real and redshift space, respectively shown in the left and middle panels, are multiplied by $k^2$ for clarity. 
 }
\label{fig:pkmu_11_lrg}
\end{figure}

Figure \ref{fig:pkmu_11_lrg} compares the results for the auto power spectra of momentum for halos and galaxies. The left panel presents the momentum auto spectrum in real space $P^{r,h}_{11}$. As we discussed in section \ref{sec:analysis_halo_p11} we assume the Poisson model for the shot noise for the momentum power spectrum as $\langle v_\parallel^2 \rangle/n$. Thus there are systematic deviations between the measurements and nonlinear PT, coming from the deviation from the Poisson model as well as the accuracy of modeling the vector part of $P_{11}^{r,h}$. 
In redshift space, as shown at the middle panel, there is a suppression of the auto power spectrum for galaxies $P^{s,h}_{11}(k,\mu)$ for $\mu=0.9$ that cannot be explained by the incorrect assumption of the Poisson model and thus can be considered as nonlinear contributions from satellite galaxies. However, like the case for the cross power shown above, $P_{11}^{s,h}$ for galaxies is relatively well modeled by nonlinear PT at one-loop level predicted for halos, compared to linear theory. 
The ratio of the redshift-space and real-space power spectra of momentum, 
$P_{11}^{s,h}(k,\mu)/P_{11}^{r,h}(k,\mu)$, is shown for halos and galaxies 
at the right panel of figure \ref{fig:pkmu_11_lrg}.
As expected, the deviation of the ratio from unity caused by nonlinear RSD is larger for galaxies that 
have contributions from satellites than for halos.

\subsection{Configuration space}
\begin{figure}[bt]
\subfigure{\includegraphics[width=.985\textwidth]{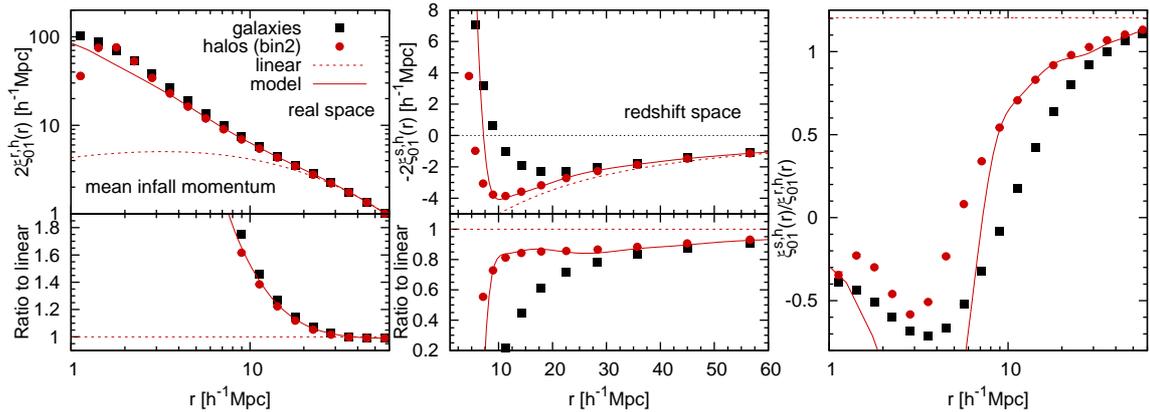}}
\caption{From the left, we show the mean pairwise infall momenta in real space, those in redshift space, and the ratios of the mean infall in redshift space and in real space. At the left panel, we show $-(-2\xi_{01}^{r,h})=2\xi_{01}^{r,h}$ because $-2\xi_{01}^{r,h}$ has a negative sign for all the scales. 
The black and red points respectively show the results for mock galaxies and halos at $z=0.5$. The halo subsample has almost the same bias as the galaxy sample. The dotted lines show the linear Kaiser model while the solid lines show our PT prediction for the halo subsample. 
 }
\label{fig:p_pair_lrg}
\end{figure}

Finally let us consider effects of satellite galaxies on velocity statistics in configuration space, the mean pairwise infall momentum and the momentum correlation function. The upper panel of the left set in figure \ref{fig:p_pair_lrg} shows the mean pairwise momentum of galaxies measured in real space at $z=0.5$. Because this statistics $-2\xi_{01}^{r,h}$ has a negative sign at all the scales, we plot $-(-2\xi_{01}^{r,h})=2\xi_{01}^{r,h}$. The lower panel shows its ratio to linear theory $\xi_{01}^{r,h}/\xi_{01,{\rm lin}}^{r,h}$. We also compare it to the measurement and one-loop nonlinear PT for halos. The measured velocity of galaxies deviates from the prediction for halos at slightly larger scales, $r\simeq 10\himpc$, but for $r>10\himpc$ the mean pairwise momentum for galaxies is almost equivalent to that for halos. While the amplitude of $\xi_{01}^{r,h}$ for halos is suppressed at small scales because of finite sizes of halos, that is not found for the galaxy sample because of the one-halo term. 

The middle set in figure \ref{fig:p_pair_lrg} is the same as the left but shows the results in redshift space $-2\xi_{01}^{s,h}$. The result for mock galaxies starts to deviate from the nonlinear PT for halos at one-loop level at $r\simeq 40\himpc$. Because of the FoG effect from satellite galaxies, the scale of upturn that appeared at $r\simeq 10\himpc$ for halos shifts to larger scales, $r\sim 20\himpc$, and is no longer predictable with the halo nonlinear PT. Thus, in order to predict the upturn scale of the mean infall velocity measured by \cite{Hand:2012} for detection of the kSZ effect, that is also at $r\sim 20\himpc$, one needs to properly take into account the effect of the FoG, that should be pursued in future work. The ratio of the mean pairwise momentum in redshift space to that in real space is shown in the top right panel in figure \ref{fig:p_pair_lrg}.

\begin{figure}[bt]
\subfigure{\includegraphics[width=.985\textwidth]{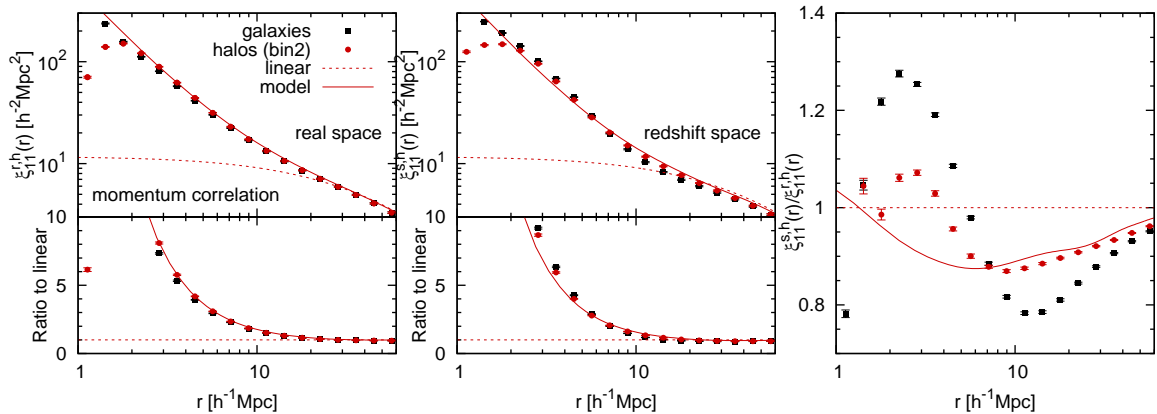}}
\caption{Same as figure \ref{fig:p_pair_lrg} but results for the momentum correlation functions are shown. 
 }
\label{fig:p_auto_lrg}
\end{figure}

Figure \ref{fig:p_auto_lrg} present the comparison of the momentum correlation function for galaxies to that for halos. The upper panel of the left set shows the real-space momentum correlation functions for 
halos and galaxies, $\xi_{11}^{r,h}$, and the lower panel shows their ratios to the linear theory prediction. 
The real-space result for galaxies is similar to that for halos, down to less than $10\himpc$. 
The middle panels in figure \ref{fig:p_auto_lrg} are the same as the left panels but 
show the results in redshift space $\xi_{11}^{s,h}$. 
The result for galaxies slightly deviates from nonlinear PT at one loop at $r\leq 30\himpc$, but the trend of 
the shape of the correlation function is still well described with the nonlinear PT down to scales similar to 
the result in real space. 
The ratio of the momentum correlation function in redshift space to that in real space is shown 
in the right panel of figure \ref{fig:p_auto_lrg}.
While the effect of nonlinear RSD on the halo momentum correlation is $\sim 10\%$ that is predicted by 
our nonlinear PT at $r\geq 5\himpc$, that on the galaxy momentum correlation has a $\sim 20\%$ effect. 
Taking into account the contribution of satellite galaxies is also an important issue 
to model a redshift-space power spectrum of galaxies. 
The effect of satellite galaxies needs to be studied for a theoretical modeling of 
velocity statistics in redshift space.


\section{Conclusions}\label{sec:conclusion}
In this paper we have studied the cross power spectrum between the density and momentum density fields together with the auto power spectrum of the momentum density field, which are the fundamental statistics being measured in peculiar velocity surveys. What we can directly measure from velocity surveys is the momentum density field in redshift space, rather than the velocity field in real space, which was the focus of most of the previous analyzes. This leads to corrections both at the linear and nonlinear level. To model them, we extended the phase-space distribution function approach originally developed to predict the redshift-space power spectrum of the density field \cite{Seljak:2011, Okumura:2012, Okumura:2012b, Vlah:2012, Vlah:2013}. In doing so, we showed that the redshift space momentum field can be expressed in terms of a sum over mass-weighted velocity moments in real space, starting from the real space momentum density field. Likewise, the momentum power spectra in redshift space can be expressed in terms of correlators between the Fourier components of these moments, analogously to the density-density power spectrum. 

We derived momentum power spectra using both the linear Kaiser model \cite{Kaiser:1987} and the perturbative approach developed in \cite{Vlah:2012, Vlah:2013}. To test the accuracy of the linear theory and phase space approach, we compared the theoretical predictions for the density-momentum cross power and momentum auto power spectra in redshift space with those measured from cosmological $N$-body simulations. We found that, for dark matter, the velocity dispersion affects the redshift-space cross power spectrum between the density and momentum. Therefore, for dark matter the numerical results deviate from linear theory at $k \sim 0.1 \hmpci$ by $\sim 50\%$ at $z=0$. Our nonlinear PT prediction, which relies on the real-space dark matter power spectrum measured from the simulations, achieves a much better agreement than linear PT. It can reproduce the $N$-body results at $z=0$ for the transverse direction $\mu\sim 0.3$ nearly perfectly up to $k \simeq 0.3\hmpci$, and for the line-of-sight direction $\mu\sim 0.9$ up to $k\simeq 0.06 \hmpci$. The accuracy of our nonlinear PT is even better for the momentum auto power spectrum, because it is less affected by the velocity dispersion. It predicts the $N$-body measurement for dark matter at $z=0$ for the line-of-sight direction $\mu\sim 0.9$ up to $k\simeq 0.08 \hmpci$. 

Because the dark matter halos have smaller peculiar velocity than the dark matter particles, the momentum power spectra for halos are better modeled than those for dark matter by combining nonlinear PT at one-loop level with a physically motivated description of halo biasing. However, the auto power spectra of the momentum field $P_{11}^{r,h}$ and $P_{11}^{s,h}$ are affected by exclusion and other nonperturbative effects that require furhter understanding in order to precisely model the momentum auto power spectrum of sparse, biased objects. 
 
We have also presented an analysis of peculiar velocity statistics in configuration space. We measured the mean infall pairwise momentum and the momentum correlation function for both dark matter and halos. We then compared these measurement in real and redshift space to our one-loop PT calculations, finding significant improvements relative to linear theory. As an example of a realistic galaxy sample, we constructed a mock BOSS-type galaxy sample by applying the HOD modeling to the simulated halos at $z=0.5$. Since the inclusion of the effect of satellite galaxies on velocity statistics is not straightforward within our PT formalism, we simply tested how well our nonlinear PT of halos can predict peculiar velocity statistics of galaxies. We found that a more detailed PT model is required to improve the agreement. 

\acknowledgments We would like to thank Tobias Baldauf and Beth Reid for useful discussions and Stephen Appleby for careful reading of this manuscript. We also thank the anonymous referee for many useful suggestions. This research was supported by the DOE, WCU grant R32-10130, Ewha Womans University research fund 1-2008-2935-001-2, and the Swiss National Foundation under contract 200021-116696/1. V.D. acknowledges support by the Swiss National Science Foundation.

\appendix
\section{Real-space velocity statistics of halos}\label{sec:real_space}
Peculiar velocity statistics were studied in real space in detail in previous works. The main purpose of this paper is to present the formalism and numerical analysis for the peculiar velocity statistics in redshift space. However, it is still useful to present the real-space velocities to compare to the redshift-space velocity. Thus in this appendix we present the real-space velocity statistics for halos. 

\begin{figure}[bt]
\includegraphics[width=.985\textwidth]{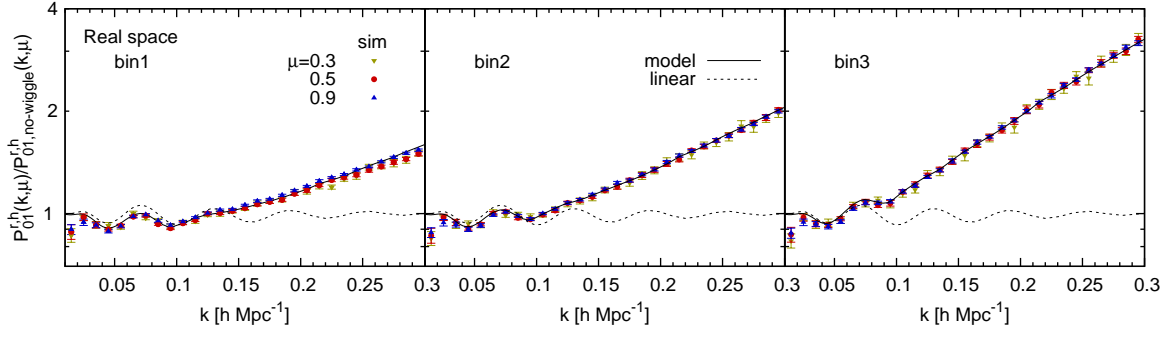}
\caption{
Density-momentum cross power spectra for halos in real space divided by the no-wiggle spectra at $z=0$, $P_{01}^{r,m}/P_{01,\text{no-wiggle}}^{r,m}$. The corresponding linear theory predictions and our model, both of which are isotropic, are shown as the dotted and solid lines, respectively.
}
\label{fig:pkmu_01_real}
\end{figure}

\begin{figure}[bt]
\includegraphics[width=.985\textwidth]{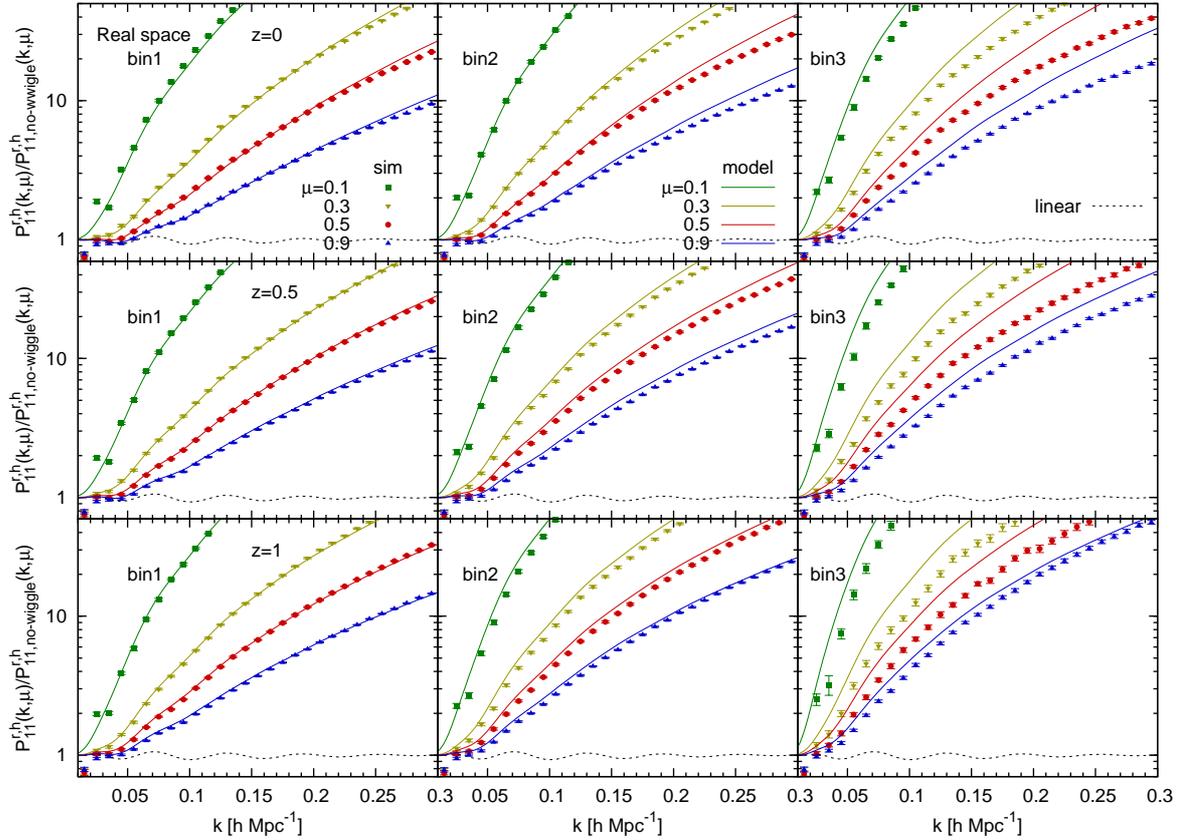}
\caption{
Momentum auto power spectra for halos in real space
 divided by the no-wiggle spectra $P_{11}^{r,h}/P_{11,\text{no-wiggle}}^{r,h}$ at $z=0$ (top row), $z=0.5$ (second row) and $z=1$ (bottom row). 
}
\label{fig:pkmu_11_real}
\end{figure}

\subsection{Fourier space}

Because we studied the density-momentum power spectrum of halos in real space in detail in \cite{Okumura:2012b, Vlah:2013}, the ratio of $P_{01}^{r,h}$ to the corresponding linear theory, $P_{01}^{r,h}/P_{01,\text{no-wiggle}}^{r,h}$ is shown only for $z=0$ in figure \ref{fig:pkmu_01_real}.
As we have seen in \cite{Okumura:2012b}, the excess of the amplitude of $P^{r,h}_{01}$ due to the nonlinear effects is larger for more massive halos. 
Since we use the best fit values of $b_1$ and $b_2$ found in \cite{Vlah:2013}, our prediction for $P_{01}^{r,h}$ agree perfectly with the $N$-body results, under the condition that the dark matter spectra $P^{r,m}_{01}$ and the scalar part of $P^{r,m}_{11}$ are perfectly modeled. 
The ratios of the real-space momentum spectra to the corresponding linear no-wiggle spectra are presented at the first, second and bottom rows in figures \ref{fig:pkmu_11_real} for $z=0$, 0.5, 1, respectively. At all the redshifts, $P^{r,h}_{11}$ is well explained by our model for less massive halos at lower redshifts. However, there are systematic deviations of $N$-body results for more massive halos from the corresponding PT due to the assumption of the Poisson shot noise. 

\subsection{Configuration space}
\begin{figure}[bt]
\includegraphics[width=.985\textwidth]{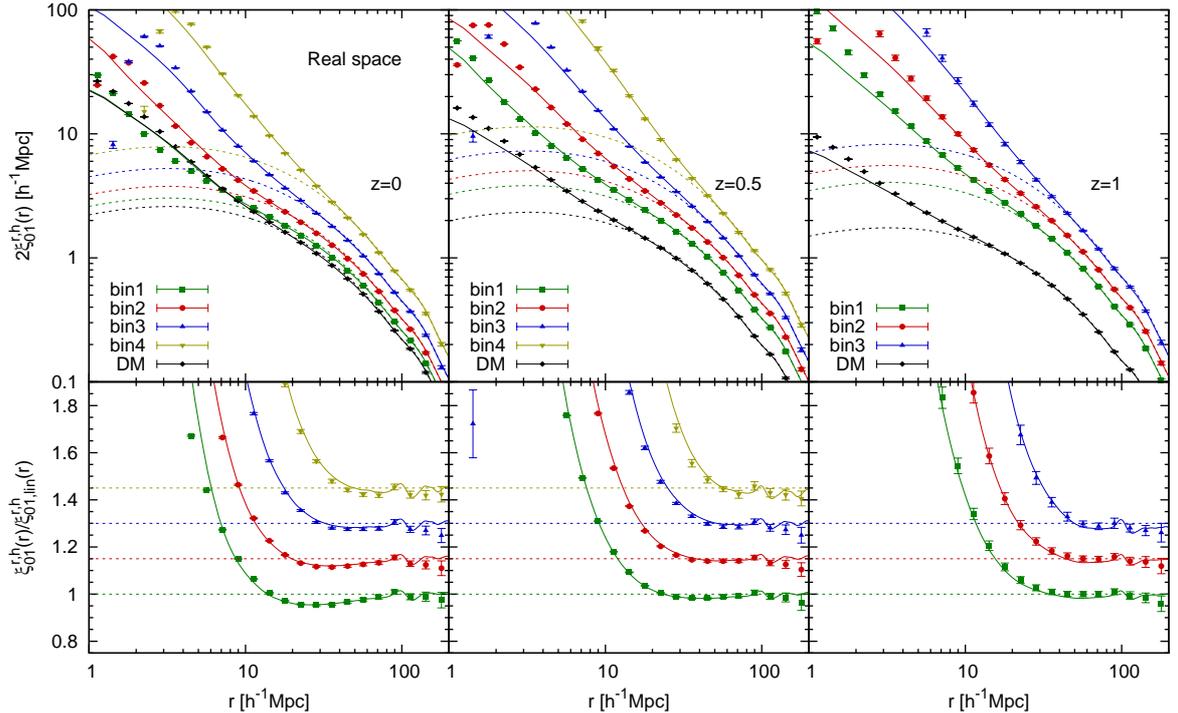}
\caption{{\it Top panels}: Mean pairwise infall momenta of halos in real space. Because $-2\xi_{01}^{r,h}$ has a negative sign at all the scales, we plot $-(-2\xi_{01}^{r,h}(r))=2\xi_{01}^{r,h}(r)$. 
  For comparison the results for dark matter are shown as the black points.
  The dotted lines with the same color are the
  corresponding linear theory predictions while the solid lines are the Fourier transform of the prediction for $2P_{01}^{r,h}$.
  {\it Bottom panels}: The ratios of the
  mean pairwise momentum to the corresponding linear theory. For
  clarity, the results are offset by $+0\%$, $+15\%$, $+30\%$ and
  $+45\%$ for halos from the lightest to heaviest. }
\label{fig:p_pair_r}
\end{figure}

The real-space mean pairwise momenta of halos are shown at the top row in figure \ref{fig:p_pair_r}, compared to those of dark matter denoted as the black points. Because the mean pairwise momenta in real space $-2\xi_{01}^{r,h}$ has a negative sign for all the scales, we plot $-(-2\xi_{01}^{r,h})=2\xi_{01}^{r,h}$ with a logarithmic scale. More massive halos have larger amplitude for the pairwise streaming momentum because it is proportional to bias. At the bottom row we show their ratios to linear theory predictions. At $z=0$, $2\xi^{r,h}_{01}$ of smaller halos starts to deviate from linear theory at larger scales, but that of larger halos diverges at larger scales, as we have seen in different statistics \cite[e.g.,][]{Okumura:2011,Reid:2011a,Okumura:2012b}. At $z>0$ the deviation from linear theory starts at larger scales for more massive halos ($\sim 40\himpc$ for bin4 halos at $z=0.5$) than for less massive halos ($\sim 20\himpc$ for bin1 halos). The solid lines are the Fourier transform of the prediction for $-2P_{01}^{r,h}$ we discussed in previous section. It agrees with the $N$-body measurements for all the four redshifts and for the scales much smaller than $r\sim 10\himpc$.

\begin{figure}[bt]
\subfigure{\includegraphics[width=.985\textwidth]{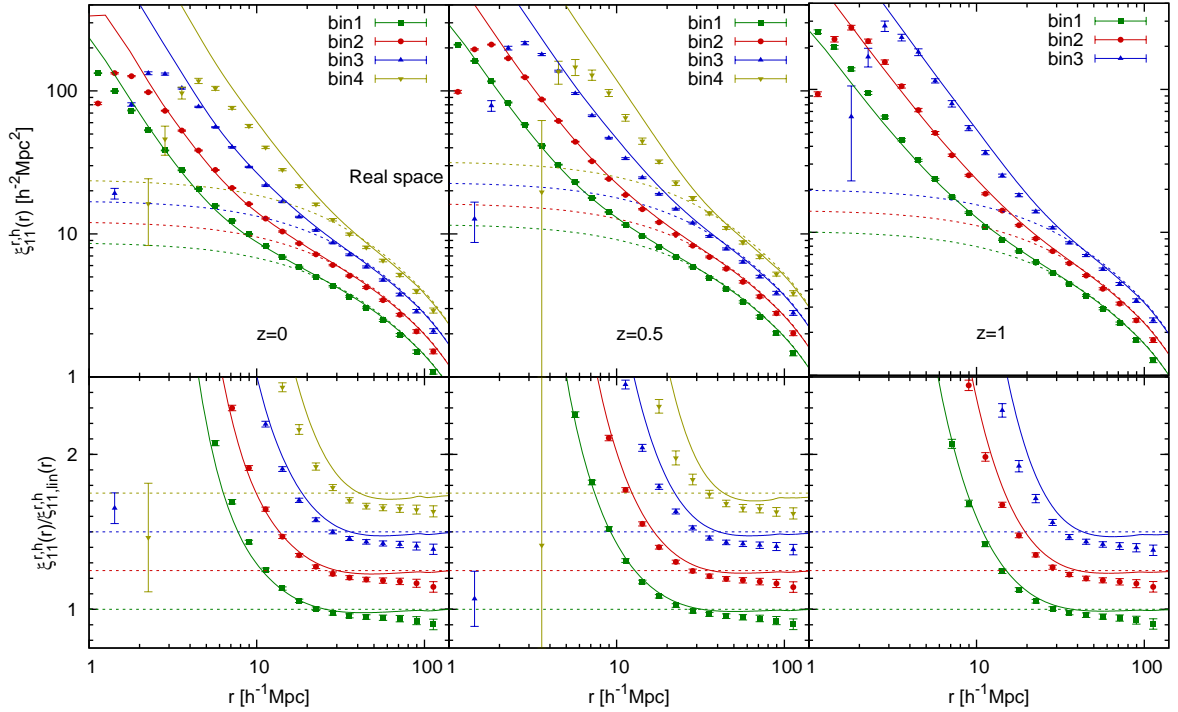}}
\caption{(Top) Momentum correlation functions of halos in real space
  $\xi^{r,h}_{11}(r)$. The dotted and solid lines show the
  corresponding linear and nonlinear PT predictions, respectively. 
  Since the correlation does not depend on bias in linear theory, the amplitude is multiplied by 
$1.4^0$, $1.4^1$, $1.4^2$ and $1.4^3$ for bin 1, 2, 3 and 4 results for clarity. 
  (Bottom) The ratios of the
  momentum correlation function to the corresponding linear theory. For
  clarity, the results are offset by $0\%$, $+25\%$, $+50\%$ and
  $+75\%$ for halos from the lightest to heavies. }
\label{fig:p_auto_r}
\end{figure}

In figure \ref{fig:p_auto_r} we present the monopole of the momentum correlation functions of halos $\xi^{r,h}_{11}(r)$
with different mass bins at redshifts $z=0$, 0.5, 1 and 2. 
Since the velocity correlation does not depend on halo bias in linear theory, we multiply the amplitude by $1.4$, $1.4^2$ and $1.4^3$ for bin 2, 3 and 4 results, respectively, for clarity at the top panels. 
$N$-body results for the momentum correlation function are well explained within linear theory 
at large scales. 
The result for more massive halos starts to deviate from linear theory at slightly larger scales, 
$\sim 30\himpc$ for the most massive halos, while $\sim 20\himpc$ for low mass halos.
Our model predicts the simulation results down to much smaller scales. However, 
because of the finite size of halos, the amplitude of the measured momentum correlation is suppressed at 
small scales.
It starts at larger scales for more massive halos that have larger size. 
As is the case for dark matter presented in figure \ref{fig:p_auto_dm}, the amplitude of the momentum correlation is systematically smaller than 
linear theory at large scales as seen in the bottom panels of figure \ref{fig:p_auto_r}. 
The trend can be seen in the work by \cite{Sheth:2009} and \cite{Bhattacharya:2008} 
that was based on linear theory and a halo model, respectively. 
Nonlinear PT improves the accuracy at the large scales, particularly for massive halos by $\sim 5\%$, 
but the measured amplitude is still systematically higher. 


\section{Density-weighted vs pair-weighted velocities}\label{sec:pair_velocity}
In this paper we focus on velocity statistics based on density-weighted velocity. It is worth comparing the accuracy of the nonlinear PT for density-weighted velocity statistics to that for pair-weighted velocity statistics, which are often used in literature \cite[e.g.,][]{Fisher:1995, Sheth:2001, Scoccimarro:2004, Hand:2012}. 

\begin{figure}[bt]
\subfigure{\includegraphics[width=.985\textwidth]{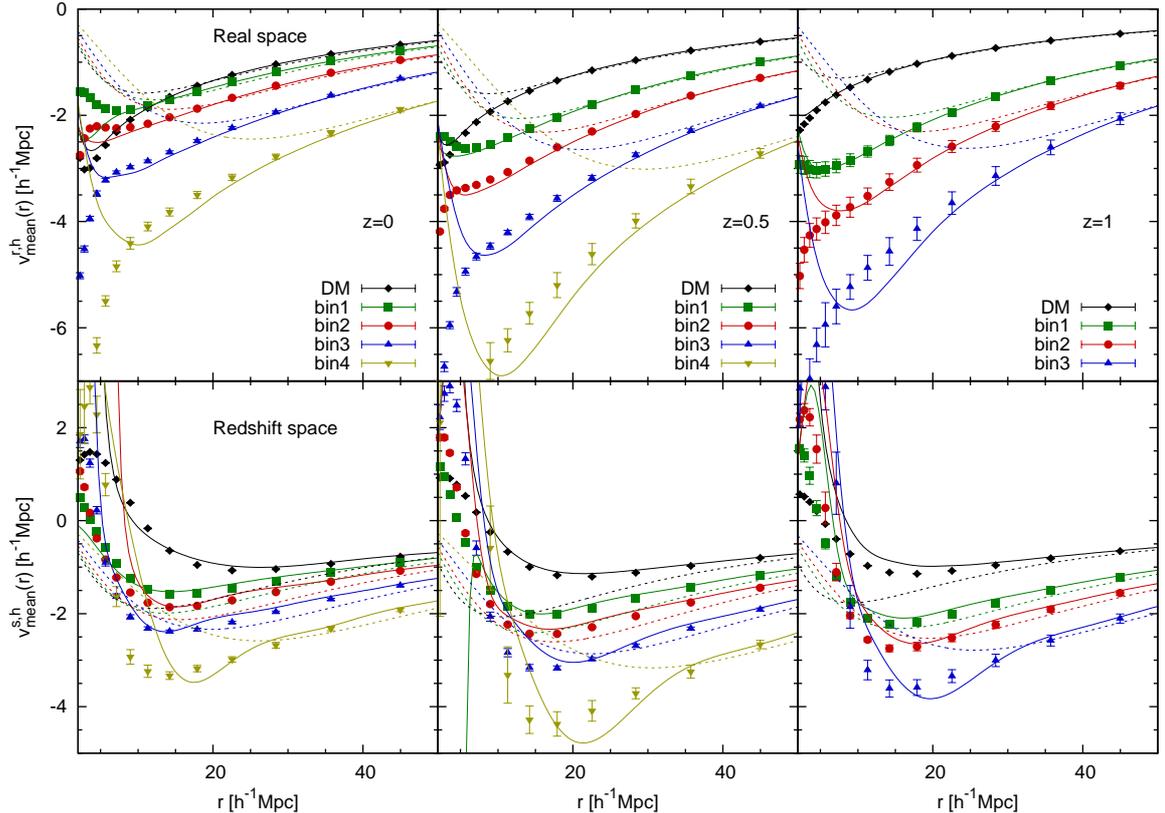}}
\caption{Pair-weighted mean pairwise velocities for dark matter and halos in real space
  $v_{\rm mean}^r(r)=\xi_{01}^r(r)[1+\xi^r(r)]^{-1}$ (top) and in redshift space
  $v_{\rm mean}^s(r)=\xi_{01}^s(r)[1+\xi^s(r)]^{-1}$ (bottom). 
  The dashed lines with the same color are the
  corresponding linear theory predictions while the solid lines are the corresponding nonlinear PT predictions. 
}
\label{fig:vol_vel01}
\end{figure}

\subsection{Mean pairwise velocity}
The pair-weighted pairwise velocity in real space $v_{\parallel{\rm mean}}^r$ and in redshift space $v_{\parallel{\rm mean}}^s$ is given by 
\be
v_{\parallel{\rm mean}}^X(\vr)\equiv 
\frac{\left\langle \left[1+\delta^X(\vx_a)\right]\left[1+\delta^X(\vx_b)\right]
\left[v_\parallel^X(\vx_b)-v_\parallel^X(\vx_a)\right] \right\rangle}
{\left\langle \left[1+\delta^X(\vx_a)\right] \left[1+\delta^X\left(\vx_b\right)\right] \right\rangle}
=
\frac{-2\xi_{01}^X(\vr)}{1+\xi^X(\vr)},  \label{eq:mean_vel_vol} 
\ee
where $\vr=\vx_b-\vx_a$.
As is clear from equation (\ref{eq:mean_vel_vol}), in order to predict the pair-weighted pairwise velocity in real and redshift space one needs to additionally model the density correlation function in real and redshift space, respectively, that comes into the denominator. The effect of the denominator becomes more significant for a sample with the higher bias. It is one of the disadvantages to use the pair-weighted velocity. Moreover, the pair-weighted velocity is not equivalent to volume-weighted velocity, namely $v^X_{\parallel{\rm mean}}\neq -2\langle \delta^X(\vx_b)v_\parallel^X(\vx_a) \rangle$.

The dipole component of the pair-weighted mean infall velocity in real space, $v_{\rm mean}^{r,h}(r)$, 
measured using the estimator of equation (\ref{eq:mean_vel_est}), is shown in the top row of figure \ref{fig:vol_vel01}. For linear theory denoted by the dashed lines, the correlation function in the denominator is also computed using linear theory $\xi^{r,h}_{\rm lin}=b_1^2 \xi^{r,m}_{\rm lin}$. The agreement of the $N$-body result to linear theory differs from the case for the density-weighted velocity by a factor of $(1+\xi^{r,h}(r))/(1+b_1^2\xi_{\rm lin}^{r,m}(r))$. At small scales where the $N$-body results largely deviate from linear theory ($r<10\himpc$), the deviation becomes smaller for the pair-weighted mean infall velocity at a glance because the nonlinearity in $\xi_{01}^r$ and that in $\xi^r$ cancel out to some extent. However, it does not mean that the pair-weighted velocity is better modeled by linear theory than density-weighted velocity.
The black solid lines are the predictions for dark mater, $v_{\rm mean}^{r,m}=-2\xi^{r,m}_{01}/[1+\xi^{r,m}]$, where $\xi^{r,m}$ is also measured from simulations, thus the accuracy is the same as the solid lines at the top panel of the left set in figure \ref{fig:p_pair_dm}.
The solid curves with other colors are the predictions for halos, $v_{\rm mean}^{r,h}$, where the measured $\xi^{r,m}$ and the nonlocal bias model are used \cite{Vlah:2013}. 
In figure \ref{fig:p_pair_r} we have seen that the measured density-weighted pairwise velocity 
for halos in real space is accurately modeled down to $r<10\himpc$. 
On the other hand, that of the pair-weighted velocity agrees with $N$-body results only at larger scales. Particularly for the most massive halo subsamples, the disagreement starts at $r\sim 30\himpc$ at $z=0$ and 0.5. 

The dipole component of the pair-weighted mean pairwise velocity in redshift space, 
$v_{\rm mean}^{s}(r)=-2\xi^s_{01}(r)[1+\xi^{s}(r)]^{-1}$,
is the statistics used by \cite{Hand:2012} to claim the detection of the kSZ effect. 
The bottom row of figure \ref{fig:vol_vel01} shows the measurements of $v_{\rm mean}^{s}$
using the estimator of equation (\ref{eq:mean_vel_est}).
Here we model the redshift-space correlation function $\xi^s$ using nonlinear PT in the same way as \cite{Vlah:2013}. 
Briefly, in equation (\ref{eq:p_ss_ang}) we use the $N$-body results for $P_{00}^{r,m}$, $P_{01}^{r,m}$ and the scalar part of $P_{11}^{r,m}$, nonlinear PT for the other terms up to $\mu^4$. Then for dark matter we use the best fit parameters for the nonlinear velocity dispersion as we did for the redshift-space momentum power spectrum $P_{01}^{s,m}$ and $P_{11}^{s,m}$, while for halos we use the simple model of the linear theory velocity dispersion and adopt the nonlocal bias model. 
The $N$-body results for the pair-weighted mean pairwise velocity are 
predicted by the nonlinear PT for both dark matter and halos, as well as those for the density-weighted velocity shown in figure \ref{fig:p_pair_s}, up to the peak scales at $\sim 20\himpc$. 

\begin{figure}[bt]
\subfigure{\includegraphics[width=.985\textwidth]{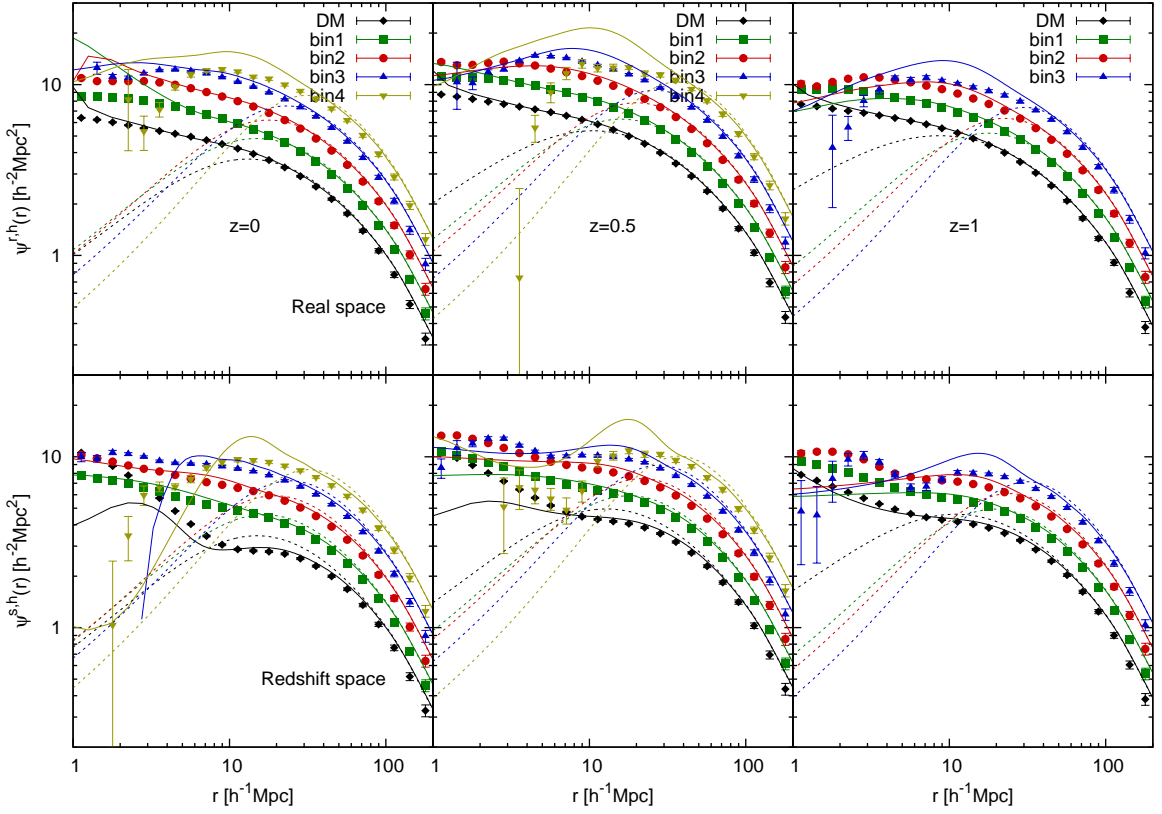}}
\caption{Same as figure \ref{fig:vol_vel01} but the pair-weighted velocity correlation functions
in real space $\psi^r(r)=\xi^r_{11}(r)[1+\xi(r)]^{-1}$ (top) 
and in redshift space $\psi^{s}(r)=\xi^{s}_{11}(r)[1+\xi^s(r)]^{-1}$ (bottom). 
The amplitude is multiplied by $1.4^{-1}$ for 
results for dark matter, and by $1.4^{0}$, $1.4^{1}$, $1.4^{2}$, and $1.4^{3}$ for halos with 
from the lightest mass (bin1) to the largest mass (bin4).}
\label{fig:vol_vel11}
\end{figure}


\subsection{Velocity correlation function}
The velocity correlation function, studied in the previous section, can also be
defined similarly as pair-weighted statistics. 
It is expressed in real space $\psi^r$ and in redshift space $\psi^s$ as
\be
	\psi^X(\vr)
	 = \frac{(1-\mu^2)\Psi_\perp^X(r)+\mu^2\Psi_\partial^X(r)}
	 {\left\langle \left[1+\delta^X(\vx_a)\right]\left[1+\delta^X(\vx_b)\right]\right\rangle}
	 =\frac{\xi^X_{11}(\vr)}{1+\xi^X(\vr)}, \label{eq:vel_corr_vol}
\ee
where $X=r$ or $s$. Compared to the density-weighted velocity correlation $\xi_{11}^X$, modeling the pair-weighted correlation $\psi^X$ requires the additional modeling of the density correlation function $\xi^X$.

First we focus on the velocity correlation measured in real space $\psi^r$. \cite{Croft:1995, Peel:2006} claimed that linear theory fails to predict the real-space velocity correlation function even at very large scales, but later it was pointed out by \cite{Sheth:2009} that the disagreement was because of the confusion of the pair-weighted and density-weighted velocity statistics. 
As shown at the top row of figure \ref{fig:vol_vel11}, the pair-weighted velocity correlation function of halos measured from simulations in real space $\psi^r$ agrees with linear theory up at $r>20\himpc$ at $z=0$ and $r>40\himpc$ at $z=1$, 
confirming the work by \cite{Sheth:2009}. This accuracy is more or less the same as the case for 
the density-weighted velocity correlation in real space shown in figure \ref{fig:p_auto_r}.
Our model predicts the $N$-body results in real space with similar accuracy to 
the case for the number-weighted velocity correlation because the real-space halo correlation function
$\xi^{r,h}$ is precisely modeled with the perfect modeling of $P^{r,m}_{00}$ and the nonlocal bias model. 

Next we present the results of the pair-weighted velocity correlation function measured in redshift space $\psi^{s}$ at the bottom row of figure \ref{fig:vol_vel11}. Note that the corresponding linear theory in redshift space is no longer the same as that in real space,
$\psi^s_{11,{\rm lin}}\ne\psi^r_{11,{\rm lin}}$ despite $\xi^s_{11,{\rm lin}}=\xi^r_{11,{\rm lin}}$ because $\xi^s_{\rm lin}=(b+f\mu^2)^2\xi^r_{\rm lin}$.
Good agreement between numerical results and one-loop nonlinear PT is found for 
both dark matter and halos for the similar 
range to the case for the number-weighted velocity correlation shown in figures \ref{fig:p_auto_dm} and 
\ref{fig:p_auto_s}.
However, at the scales lower than the valid range of nonlinear PT, the disagreement in the 
halo correlation becomes 
large very quickly because of large nonlinearity of halo clustering $\xi^{s,h}$. 
Also note again that the pair-weighted velocity correlation function defined by equation (\ref{eq:vel_corr_vol}) is not equivalent to the volume-weighted velocity correlation function $\psi^X(\vr)\neq \langle v_\parallel^X(\vx_b) v_\parallel^X(\vx_a) \rangle$.
We thus conclude that there is no advantage in defining pair-weighted velocity statistics that are commonly used in previous works.

\bibliography{ms.bbl}
\bibliographystyle{revtex}

\end{document}